\pdfoutput=1
\documentclass[aps,prd,reprint,superscriptaddress,longbibliography,nofootinbib]{revtex4-1}

\usepackage{graphicx}
\usepackage{amsmath}
\usepackage{amssymb}
\usepackage{amsfonts}
\usepackage{dcolumn}
\usepackage{dsfont}
\usepackage{latexsym}
\usepackage{rotating}
\usepackage{color}
\usepackage{latexsym}
\usepackage{bbm}
\usepackage{subfigure}
\usepackage{float}
\usepackage{epsfig}
\usepackage{psfrag}
\usepackage{natbib, hyperref}
\usepackage{bm}
\usepackage{amsthm}
\usepackage{eucal}
\usepackage{mathrsfs}
\usepackage{url}
\usepackage{braket}
\usepackage{ulem}
\usepackage{calligra}
\usepackage[T1]{fontenc}
\usepackage[utf8]{inputenc}
\usepackage{newunicodechar}
\usepackage{marvosym}
\usepackage[absolute]{textpos}
\usepackage[cal=cm]{mathalfa}
\usepackage{soul}
\usepackage{amsmath}
\usepackage{bbold}

\usepackage{color}

\hypersetup{
colorlinks=true,final=true,
        linkcolor=blue,
        citecolor=blue,
        filecolor=blue,
        urlcolor=blue,
}

\begin{document}
\setlength{\abovedisplayskip}{3pt}
\setlength{\belowdisplayskip}{3pt}

\title{Semi-Dirac and Weyl Fermions in Transition Metal Oxides}

\author{Narayan Mohanta}
\affiliation{Materials Science and Technology Division, Oak Ridge National Laboratory, Oak Ridge, TN 37831, USA}
\author{Jong Mok Ok}
\affiliation{Materials Science and Technology Division, Oak Ridge National Laboratory, Oak Ridge, TN 37831, USA}
\author{Jie Zhang}
\affiliation{Materials Science and Technology Division, Oak Ridge National Laboratory, Oak Ridge, TN 37831, USA}
\author{Hu Miao}
\affiliation{Materials Science and Technology Division, Oak Ridge National Laboratory, Oak Ridge, TN 37831, USA}
\author{Elbio Dagotto}
\affiliation{Materials Science and Technology Division, Oak Ridge National Laboratory, Oak Ridge, TN 37831, USA}
\affiliation{Department of Physics and Astronomy, The University of Tennessee, Knoxville, TN 37996, USA}
\author{Ho Nyung Lee}
\affiliation{Materials Science and Technology Division, Oak Ridge National Laboratory, Oak Ridge, TN 37831, USA}
\author{Satoshi Okamoto}
\affiliation{Materials Science and Technology Division, Oak Ridge National Laboratory, Oak Ridge, TN 37831, USA}

\begin{abstract}
We show that a class of compounds with \textit{I}4/\textit{mcm} crystalline symmetry hosts three-dimensional semi-Dirac fermions. Unlike the known two-dimensional semi-Dirac points, the degeneracy of these three-dimensional  semi-Dirac points is not lifted by spin-orbit coupling due to the protection by a nonsymmorphic symmetry---mirror reflection in the \textit{a-b} plane and a translation along the \textit{c} axis. This crystalline symmetry is found in tetragonal perovskite oxides, realizable in thin films by epitaxial strain that results in a$^0$a$^0$c$^-$-type octahedral rotation. Interestingly, with broken time-reversal symmetry, two pairs of Weyl points emerge from the semi-Dirac points within the Brillouin zone, and an additional lattice distortion leads to enhanced intrinsic anomalous Hall effect. The ability to tune the Berry phase by epitaxial strain can be useful in novel oxide-based electronic devices.
\end{abstract}
           
\maketitle

Protected band degeneracy continues being a fundamental topic of interest to understand the physics of various topological semimetals, \textit{e.g.} Dirac, Weyl, and nodal loop semimetals~\cite{Herring_PhyRev1937,Burkov_PRB2011,Burkov_NMat2016,Bzdusek_Nature2016,Amritage_RMP2018,Burkov_ARCMP2018}. The guiding principle to classify these topological semimetals has been the presence of one or more symmetries that allow degenerate eigenstates~\cite{Chiu_RMP2016,Yang_AdvPhys2018}. Spin-orbit coupling (SOC) lifts the degenerate eigenstates everywhere except at momenta where symmetry enforces that the matrix elements of the SOC operator vanish~\cite{Allen_PhysicaC2018}. Since SOC is present in all materials, the implication is that unless there is a special symmetry that protects the band degeneracy, the topological nodal semimetals exist in the theoretically-idealized case when SOC is zero.

Semi-Dirac fermions in two-dimensional solids connect the physics of monolayer graphene that exhibits linear band dispersion near the Dirac points and bilayer graphene that has quadratic band dispersion near the band-touching points~\cite{Neto_RMP2009,McCann_RPP2013}. The semi-Dirac points, in which both linear and quadratic band dispersions occur along different momentum directions, exist in the absence of SOC, in various systems including hexagonal lattices under a magnetic field~\cite{Dietl_PRL2008}, VO$_2$-TiO$_2$ heterostructures~\cite{Pardo_PRL2009,Banerjee_PRL2009}, BEDT-TTF$_2$I$_3$ salt under pressure~\cite{Katayama_JPSJ2006}, silicene oxide~\cite{Zhong_PhysChemChemPhys2017}, photonic crystals~\cite{Wu_OPTEx2014}, and Hofstadter spectrum~\cite{Delplace_PRB2010}. The quadratic dispersion along one of the momentum directions leads to an enhanced density of states at low energies as compared to linear Dirac dispersion, making instabilities relatively easier to occur due to the enlarged phase space for quantum fluctuations~\cite{Uryszek_PRB2019}. A semi-Dirac semimetal becomes a Chern insulator, revealing quantized anomalous Hall conductivity with broken time-reversal symmetry (TRS), in the presence of SOC that opens an energy gap at the semi-Dirac point~\cite{Huang_PRB2015}. A similar Chern insulating state is also induced in a semi-Dirac semimetal with two gapless Dirac nodes by impinging a circularly polarized light~\cite{Kush_PRB2016}. In multilayer (TiO$_2$)$_m$$\slash$(VO$_2$)$_n$ heterostructures, the semi-Dirac point is not destroyed by SOC in a special situation when the spins are aligned along the rutile $c$ axis~\cite{Padro_PRB2010}. 

However, the questions that remain open are whether there exists a natural material in which the semi-Dirac point is gapless generically when SOC is present and whether these semi-Dirac points exist in a higher dimension. Here, we show that a class of materials hosts three-dimensional (3D) gapless semi-Dirac fermions  in the presence of SOC. Tetragonal perovskite oxides with \textit{I}4/\textit{mcm} crystalline symmetry (space group 140) exhibit such 3D semi-Dirac points at a high-symmetry point, and its nonsymmorphic symmetry protects the semi-Dirac points against the gap-opening by SOC. Nonsymmorphic symmetry-protected two-dimensional Dirac semimetallic phase was proposed in a theoretical model~\cite{Young_PRL2015}, and subsequent studies using \textit{ab initio} calculations in layered compounds~\cite{Li_PRB2018,Shao_SciRep2018}. Cubic perovskite oxide thin films can be transformed into the tetragonal phase to realize the \textit{I}4/\textit{mcm} space group under epitaxial strain that produces a$^0$a$^0$c$^-$-type (in the Glazer notation) octahedral rotation~\cite{Rondinelli_PRB2010,JongMok_SciAdv2021}. The out-of-phase octahedral rotation brings in the nonsymmorphic symmetry: mirror reflection about $x=0$/$y=0$ plane, translation along [001] $z\rightarrow z+1/2$, as shown in Fig.~\ref{fig1}(a) (for a symmetry analysis, see the Supplemental Materials~\cite{SM} and Ref.~[\onlinecite{Hirschmann_PRMat2021}]). A consequence of the fractional translational symmetry is that the unit cell of the original symmorphic crystal is doubled. In this case, the folded back bands stick together necessarily at the Brillouin-zone boundary if the nonsymmorphic symmetry is present, even with SOC.

\begin{figure}[t]
\begin{center}
\vspace{-0mm}
\epsfig{file=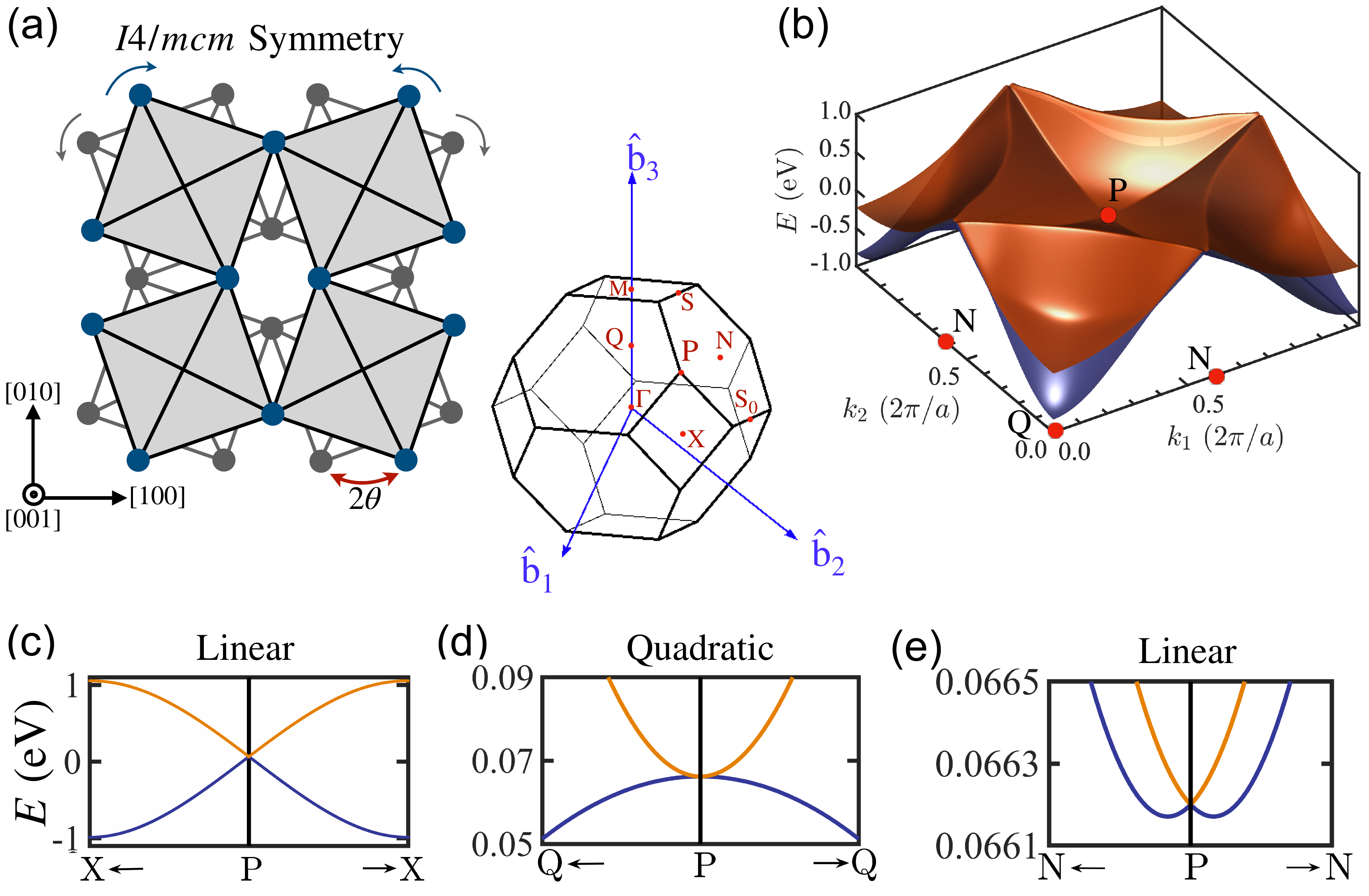,trim=0.0in 0.0in 0.0in 0.0in,clip=false, width=86mm}
\caption{(a) Illustration of the octahedral rotation-driven tetragonal phase of perovskite oxides. a$^0$a$^0$c$^-$-type rotation in a cubic crystal produces \textit{I}4/\textit{mcm} space group that exhibits the nonsymmorphic symmetry (mirror reflection in the $a$-$b$ plane and translation along the $c$ axis). (b)  Dispersion of two electronic bands in the $\rm P$-$\rm Q$-$\rm N$ plane. (c)-(e) show the dispersion curves along different momentum paths. The inset in the middle shows the high-symmetry points in the Brillouin zone.}
\label{fig1}
\vspace{-5mm}
\end{center}
\end{figure}
As a particular physical realization of the {\it I}4/{\it mcm} space group, we primarily focus on tetragonal SrNbO$_3$ that manifestly displays the defining characteristics of this nonsymmorphic symmetry, and a 3D semi-Dirac point appears close to the Fermi energy at a high-symmetry point P at the Brillouin-zone boundary. Three different kinds of dispersion emerge from this semi-Dirac point---linear along the P-X direction, quadratic along the P-Q direction, and linear in the close vicinity of the P point along the P-N direction, as shown in Fig.~\ref{fig1}(b)-(e). With a TRS-breaking element such as an external magnetic field, two pairs of Weyl points appear within the Brillouin zone; each pair is shared by the opposite zone boundaries. The Weyl points in semi-Dirac semimetals lead to unique properties such as highly-anisotropic plasmon frequency, unusual carrier density-dependence of the Hall coefficient, and the divergence of the ratio of orbital to spin susceptibilities at low carrier doping~\cite{Banerjee_PRB2012}.  The Weyl points in tetragonal SrNbO$_3$ open up an energy gap with an additional broken inversion symmetry, obtained by a small displacement at a Nb site. This gapped Weyl points, at the $\rm P$ point, exhibits an enhanced intrinsic anomalous Hall effect. With these intriguing properties, perovskite oxides of the \textit{I}4/\textit{mcm} space group provide a unique example of a class of compounds that hosts crystalline symmetry-protected 3D semi-Dirac fermions. Our Berry curvature analysis sheds light on the unusual features in transport experiments, such as the Shubnikov de Haas oscillations and anomalous Hall effect measurements. Furthermore, the ability to control the Berry phase by octahedral rotation shows a pathway towards developing novel electronic devices, including high-frequency rectifiers~\cite{Isobe_SciAdv2020,Zhang2021}.

\begin{figure}[b]
\begin{center}
\vspace{0mm}
\epsfig{file=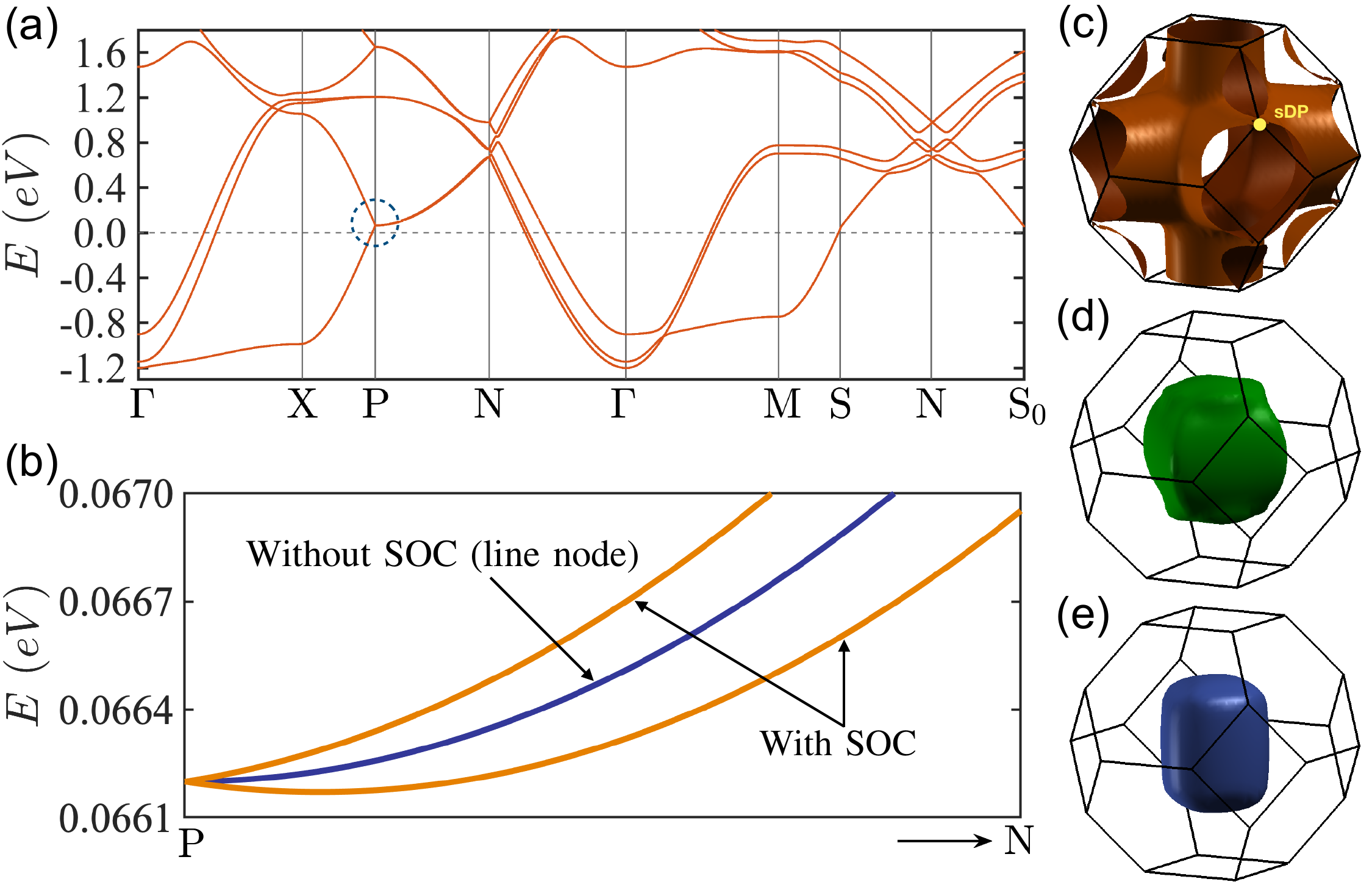,trim=0.0in 0.0in 0.0in 0.0in,clip=false, width=88mm}
\caption{(a) Band dispersion of tetragonal SrNbO$_3$ in the presence of spin-orbit coupling, through various high-symmetry momenta in the Brillouin zone. (b) Expanded view of the dispersion along the P-N direction, without and with spin-orbit coupling. (c)-(e) The three Fermi surfaces of the spin-orbit coupled $t_{2g}$ bands at the Fermi level. The semi-Dirac point (sDP) is denoted by the dashed circle at the P point in (a) and by the yellow dot in (c).}
\label{fig2}
\vspace{-5mm}
\end{center}
\end{figure}

\noindent {\bf \\Electronic structure:} The \textit{ab initio} band dispersion of tetragonal SrNbO$_3$ with a finite SOC along high-symmetry momenta is shown in Fig.~\ref{fig2}(a)~\cite{SM}. The $t_{2g}$ orbitals of Nb $4d$ state predominantly populate the Fermi level, as in the case of cubic SrNbO$_3$~\cite{Bigi_PRM2020,Park_CommunPhys2020}. Due to the zone folding, the bands along the $\rm P$-$\rm N$ momentum direction are four-fold degenerate without SOC, creating line nodes. The degeneracy of these nodal lines is lifted everywhere, except at the high symmetry points, when SOC is turned on, as shown in Fig.~\ref{fig2}(b). The TRS and the nonsymmorphic symmetry protect the semi-Dirac degeneracy at the $\rm P$ point at $E\! \approx \!0.0662$~eV. Three other Dirac points appear at the $\rm N$ point above $E\!=\!0.5$~eV. 
Being away from the Fermi level, the Dirac points at the $\rm N$ point may not be accessible to transport; here we, therefore, focus on the semi-Dirac point at the $\rm P$ point. Yet, replacing Nb with other transition-metal element such as Mo can tune the Fermi level close to the Dirac points at the N point. The Fermi surfaces of the spin-orbit coupled $t_{2g}$ orbitals are depicted in Fig.~\ref{fig2}(c)-(e). Due to the linear dispersion near the semi-Dirac point, a small chemical doping or gating can tune the Fermi level to the semi-Dirac point, creating the possibility to obtain a quantized Berry phase and a very high-mobility conduction~\cite{Wang_PRB2011,Liang_NMat2015,Veit_NatComm2018}. Also, the presence of the semi-Dirac points in the asymmetric, convex Fermi surface can produce a large nonsaturating magnetoresistance~\cite{He_NatCommun2019,Vashist_PRB2019,Zhang_PRB2019}.

\begin{figure}[t]
\begin{center}
\epsfig{file=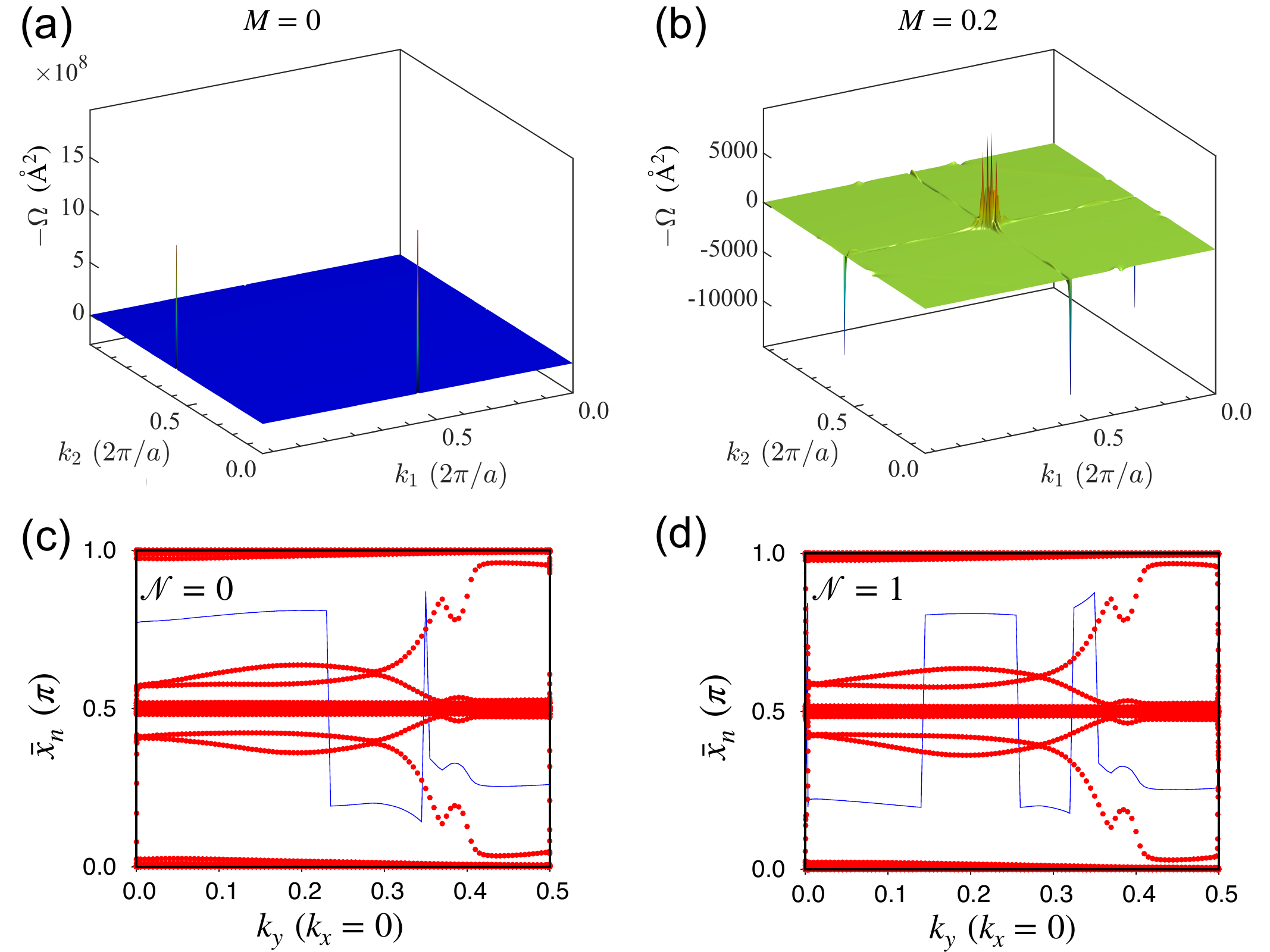,trim=0.0in 0.0in 0.0in 0.0in,clip=false, width=89mm}
\caption{Berry curvature $\Omega(\mathbf{k})$ for Nb spin polarization (a) $M\!=\!0$, and (b) $M\!=\!0.2$, plotted in a momentum plane containing the $\rm P$ point at the center, and the $\rm N$ points at the side centers. We use a dense 100$\times$100 momentum grid. (c), (d) The Wannier charge centers $\bar{x}_n$ along momentum $k_y$ with $k_x\!=\!0$ for (a) $M\!=\!0$, and (b) $M\!=\!0.2$. The blue line represents the center of the maximum gap between the Wannier charge centers. In (d), the gap center crosses the Wannier charge centers an odd number of times, resulting in a topological invariant ${\cal N}\!=\! 1$.}
\label{fig3}
\vspace{-7mm}
\end{center}
\end{figure}

\noindent {\bf \\Topological invariant:} The topological properties of tetragonal SrNbO$_3$ are investigated by two protocols---first, by looking at the Berry curvature, and second, by computing the topological invariant of the Bloch bands. The Berry curvature is given by~\cite{Wang_PRB2006}
\begin{align}
\Omega(\mathbf{k})\!=\!-\sum_{m, n}^{\varepsilon_{{m}\mathbf{k}} \neq \varepsilon_{{n}\mathbf{k}}} 2~{\rm Im}\frac{\langle \psi_{n\mathbf{k}}| v_x | \psi_{m\mathbf{k}} \rangle \langle \psi_{m\mathbf{k}}| v_y | \psi_{n\mathbf{k}} \rangle}{ (\varepsilon_{{m}\mathbf{k}}-\varepsilon_{{n}\mathbf{k}})^2},
\end{align}
where $|\psi_{n\mathbf{k}} \rangle$ is the spinor Bloch wave function of the $n^{\rm th}$ band with eigenenergy $\varepsilon_{{n}\mathbf{k}}$, and $\mathbf{v} \!=\!(v_x, v_y)$ is the velocity operator. We use Wannier interpolation based on maximally-localized Wannier functions~\cite{Marzari_PRB1997,Souza_PRb2001}. The Berry curvature in the $\rm P \!-\! N \!-\! Q$ plane is shown in Fig.~\ref{fig3}(a). The semi-Dirac point, being gapless, reveals zero Berry curvature at the center $\rm P$ point. We introduce TRS breaking and opening of an energy gap at the semi-Dirac point by constraining to a spin polarization at the Nb sites. Fig.~\ref{fig3}(b) shows a large Berry curvature contributions at and near the $\rm P$ point with $M\!=\!0.2$, which represents 20$\%$ spin polarization along the $+z$ direction. $M$ is defined as $M=\sum_{a\in {\rm Nb}~t_{2g} {\rm~orbitals}} \langle n_{a,\uparrow} \rangle -\langle n_{a,\downarrow} \rangle$, where $n_{a,\uparrow}$ is the carrier density in orbital $a$ and spin $\uparrow$; the magnetization is compatible with the magnetic space group $4/mm’m’$. The enhanced Berry curvature induced by a nonzero $M$ strongly suggests the nontrivial band topology. To check this possibility, we compute the topological invariant via the Wannier charge centers (WCCs) $\bar{x}_n$, corresponding to the Wannier functions that are maximally localized in one dimension, defined as~\cite{Soluyanov_PRB2011}
\begin{align}
\bar{x}_n\!=\! \frac{i}{2\pi}\int_{-\pi}^{\pi}d\mathbf{k} \langle u_{{n}\mathbf{k}} | \partial_{\mathbf{k}} | u_{{n}\mathbf{k}} \rangle,
\end{align}
\begin{figure}[b]
\begin{center}
\epsfig{file=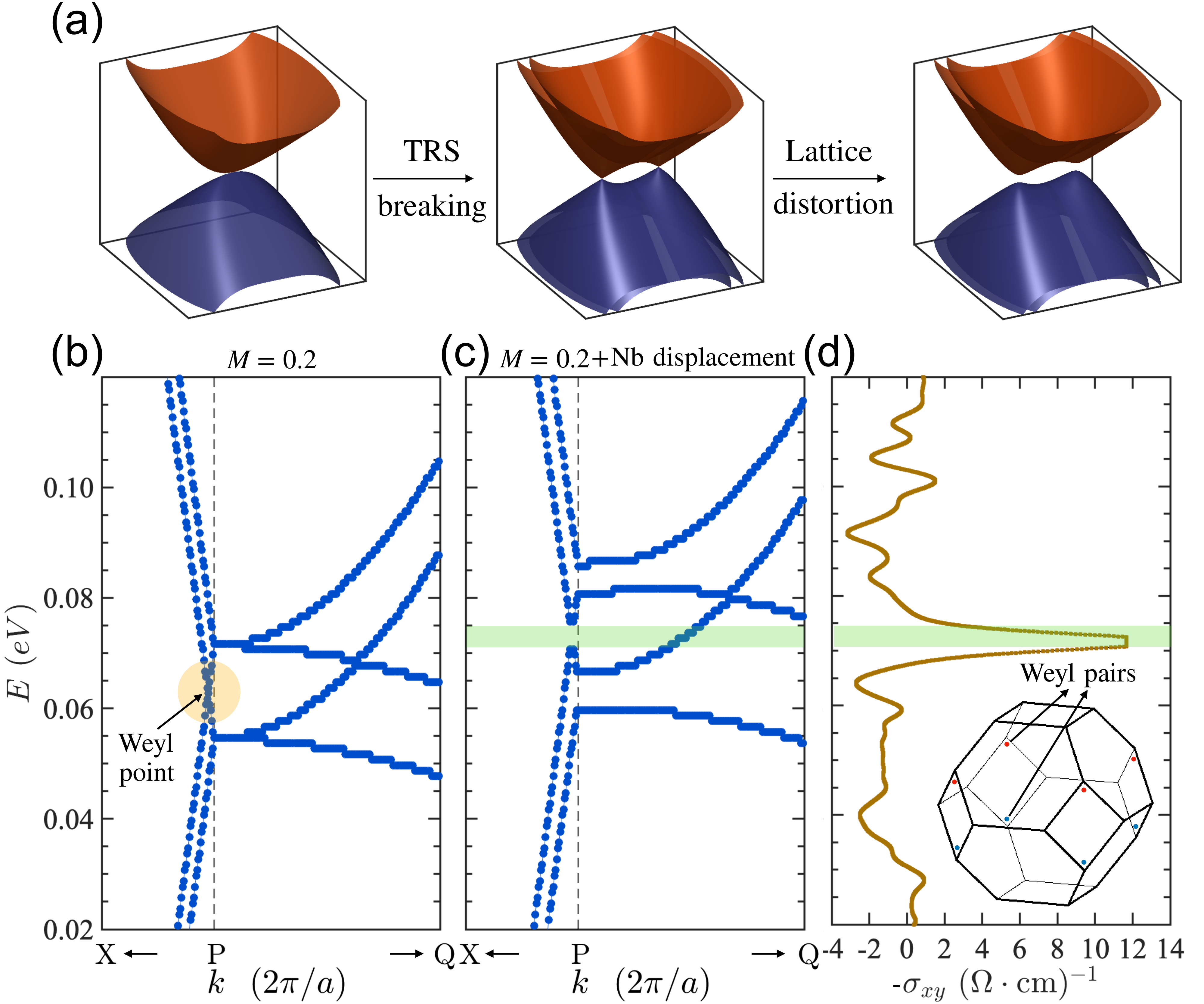,trim=0.0in 0.0in 0.0in 0.0in,clip=false, width=86mm}
\caption{(a) Effect of broken symmetries on the semi-Dirac points: TRS breaking splits the semi-Dirac point into a pair of Weyl points, and additional lattice distortion opens up an energy gap. (b) Band dispersion in the vicinity of the P point, with broken time-reversal symmetry that lifts the four-fold degeneracy of the semi-Dirac point, creating a pair of Weyl points near the P point. (c) The Weyl point acquires an energy gap with an inversion symmetry-breaking lattice distortion. (d) The anomalous Hall conductivity, with varying energy, reveals a plateau within the energy gap at the Weyl point. Inset in (d) shows the two pairs of the Weyl points in the Brillouin zone.}
\label{fig4}
\vspace{-7mm}
\end{center}
\end{figure}
\noindent where $|u_{{n}\mathbf{k}} \rangle \!=\! e^{-i\mathbf{k} \cdot \mathbf{r}} | \psi_{n\mathbf{k}} \rangle$ represents the periodic parts of the Block states. The topological invariant, originally defined in Ref.~\cite{FuKane_PRB2006}, can be computed  from the flow of the WCCs and the largest gap between the WCCs~\cite{Soluyanov_PRB2011}. In Fig.~\ref{fig3}(c) and \ref{fig3}(d), we show the evolution of the WCCs with the momentum $k_y$ for, respectively, $M\!=\!0$, and $M\!=\!0.2$. With $M\!=\!0.2$, the center of the largest gap between the WCCs (blue line) crosses the WCCs odd number of times in traversing a path from $k_y\!=\!0$ to $k_y\!=\! \pi/a$, yielding a topological invariant ${\cal N}\!=\! 1$. A non-zero ${\cal N}$ in a Chern insulator with broken  TRS usually predicts a quantum anomalous Hall effect. We, however, find that the Hall conductivity is not quantized, and the semi-Dirac points can, therefore, be classified as type I~\cite{Huang_PRB2015}, similar to those proposed in VO$_2$-TiO$_2$ heterostructures~\cite{Banerjee_PRL2009}. This is also verified by the absence of a band crossing of the surface states~\cite{SM}. The absence of topological surface states also indicates the possibility of a fragile topological phase; it, however, requires a careful analysis of appropriate symmetry indicators~\cite{Elcoro_arXiv2020,Weider_arXiv2018}.

\noindent {\bf \\Anomalous Hall effect:} In the absence of TRS, \textit{e.g.} with an external magnetic field, each semi-Dirac point splits into two Weyl points, as illustrated in Fig.~\ref{fig4}(a). In the considered tetragonal SrNbO$_3$, we find that a finite magnetization at the Nb sites generates a Weyl point close to the P point, as shown in Fig.~\ref{fig4}(b). The pair of Weyl points appears along the P-X-P momentum direction, totaling two pairs in the Brillouin zone, as shown in the inset of Fig.~\ref{fig4}(d). The Weyl points appear when the magnetization is perpendicular to the $a$-$b$ plane. Otherwise, the semi-Dirac points are directly gapped since there is no net chiral charge. By lowering the lattice symmetry further, an energy gap is induced at the Weyl points. We consider a lattice distortion that lowers the inversion symmetry but preserves the nonsymmorphic symmetry. This is achieved by shifting the center Nb site vertically by a small amount. Such a distortion may be realized naturally by epitaxial strain induced by lattice mismatch with the  substrate at a polar interface, or by applying an electric field  perpendicular to the interface. Fig.~\ref{fig4}(c) shows a distortion-induced energy gap at a Weyl point. Remarkably, within this energy gap, the intrinsic anomalous Hall conductivity acquires a plateau with an enhanced value, as shown in Fig.~\ref{fig4}(d). We also find that the anomalous Hall conductivity is dependent on the amount of distortion in the tetragonal lattice and the octahedral rotation, suggesting a route to obtain tunable anomalous Hall effect in oxide interfaces. 
  
\noindent {\bf \\Tunable Berry phase:} Understanding the Berry phase-driven phenomena in perovskite oxides is fundamentally important, not only to realize emergent electromagnetism in a feasible way, but also to identify the origin of non-trivial Hall signatures. In oxide compounds where magnetism coexists with a non-trivial band topology, such as SrRuO$_3$, the intrinsic anomalous Hall effect appears without any external magnetic field~\cite{Caviglia_PRResearch2020}. In the non-magnetic tetragonal perovskites, like the one considered here \textit{i.e.} SrNbO$_3$, doping magnetic impurities or applying an external magnetic field leads to a finite Berry phase in the Bloch bands. We find that the Berry curvature $\Omega$ can be tuned externally by changing the octahedral rotation angle $\theta$~\cite{SM}. Particularly, the Berry curvature and the resulting anomalous Hall conductivity vary non-monotonically with increasing $\theta$. Previous experiments have demonstrated that external strain can control octahedral rotations in oxide thin films~\cite{Herklotz_SciRep2016,Choquette_PRB2016}. Our results, therefore, show an effective way to control the Berry phase and  the topological properties of perovskite transition-metal oxides by epitaxial strain.

\noindent {\bf \\Effective low-energy model:} To study the low-energy behavior of the 3D semi-Dirac dispersion, we derive an effective tight-binding model for the SOC-coupled $t_{2g}$ electrons in tetragonal perovskite oxides~\cite{SM}. From this tight-binding Hamiltonian, we obtain the following low-energy Hamiltonian
\begin{align}
H_{low-E}=& 
-2 A \cos k_x \cos k_y \tau_0 \sigma_0  \nonumber \\
&+\frac{1}{2} B \tau_2  (\sin 2 k_x  \sigma_1 - \sin 2 k_y \sigma_2) \sin k_z \nonumber \\
&-2 \bigl\{C (\cos k_x + \cos k_y)  + D \cos k_z  \bigr\} \tau_1 \sigma_0 \nonumber \\
&+ 2 E  (\cos k_x + \cos k_y)  \tau_2 \sigma_3 ,
\label{lowE}
\end{align}
where $\sigma_i$ and $\tau_i$ are the Pauli matrices in the spin and the sublattice space, respectively; $A$, $B$, $C$, $D$, and $E$ are parameters. The symmetry generators and their representations in the sublattice-spin space are $E = \tau^0 \sigma^0$, $C_{2z} = i\tau^0 \sigma^z$, $IC_{4z} = \tau^0 (1-i\sigma^z)/\sqrt{2}$, $IC_{4\bar{z}} = \tau^0 (1+i\sigma^z)/\sqrt{2}$, $C_{2\bar{y}}|z = i\tau^x \sigma^y$, $C_{2\bar{x}}|z = i\tau^x \sigma^x$, $M_{xy}|z = i\tau^x (\sigma^x+\sigma_y)/\sqrt{2}$, and $M_{x\bar{y}}|z = i\tau^x (\sigma^x-\sigma_y)/\sqrt{2}$.

The partial density of states surrounding the semi-Dirac point, exhibits a depletion at the energy at which the semi-Dirac point appears~\cite{SM}, rather than a minimum as in the isotropic Dirac or 2D semi-Dirac cases~\cite{Chen_PRB2021}. A depletion of the density of states near the Fermi energy usually indicates a strong effect of electronic correlation or disorder~\cite{Noh_SSC2000}. The mutual interplay of these additional effects in the semi-Dirac compound can, therefore, produce unconventional properties.

\noindent {\bf \\Discussion:} We further analyzed the electronic properties of four other compounds from the \textit{I}4/\textit{mcm} space group, \textit{viz.} CaNbO$_3$, SrRuO$_3$, SrMoO$_3$, and SrTiO$_3$ and confirmed the presence of the 3D semi-Dirac fermions at the P point~\cite{SM}. This establishes that the discussed symmetry-protected band topology is generic to tetragonal perovskite oxides that belong to this space group, the only difference being the location in energy of the semi-Dirac points. For CaNbO$_3$ and SrRuO$_3$, the semi-Dirac point at the P point is closer to the Fermi level than in SrNbO$_3$. On the other hand, SrTiO$_3$, having a large energy gap at the Fermi level, the semi-Dirac points are situated far away from the Fermi level. Using SrTiO$_3$ as a substrate is, therefore, beneficial for providing epitaxial strain. For SrMoO$_3$ and SrRuO$_3$, three Dirac points at the N point are also located close to the Fermi level which indicates that these two compounds can be interesting also to explore the non-trivial band topology. 

In contrast to the isotropic Dirac semimetals, the semi-Dirac semimetals can be realized at a quantum critical point between a 3D Dirac semimetal and a topologically trivial insulator, at which both linear and non-linear band dispersions appear~\cite{Yang_NatCommun2014,Parameswaran_NatPhys2013}. Circularly polarized light usually does not gap the semi-Dirac nodes in rotationally-invariant systems, unlike the isotropic Dirac nodes~\cite{Narayan_PRB2015,Fiete_PRB2018,Islam_PRB2018}. Tetragonal perovskite oxides, therefore, offer a feasible material platform to study new phenomena that cannot be found in known Dirac semimetals.

The interplay of Coulomb interaction and disorder in a two-dimensional semi-Dirac semimetal produces a variety of quantum phase transitions and non-Fermi liquid behaviors~\cite{Zhao_PRB2016}. Disorder, particularly, has a profound effect as it can drive a Lifshitz transition from an insulator to a semimetal, and a topological transition to a Chern insulating state~\cite{Sriluckshmy_PRB2018}. The interplay of topology, disorder and Coulomb interaction in the nonsymmorphic symmetry-protected semi-Dirac fermions, found in the tetragonal perovskite oxides, can lead to even richer physical properties due to their three-dimensional nature. 

To conclude, we have shown that non-magnetic tetragonal perovskite oxides with \textit{I}4/\textit{mcm}  symmetry, \textit{e.g.} SrNbO$_3$, CaNbO$_3$, and SrMoO$_3$, host 3D semi-Dirac fermions that are protected by a nonsymmorphic symmetry. This crystalline symmetry can be realized by a substrate strain when the cubic perovskites undergo an octahedral rotation of type a$^0$a$^0$c$^-$. Due to the symmetry protection, the semi-Dirac degeneracy is maintained in the presence of SOC, making this class of compounds a unique, natural example. Breaking the TRS leads to two pairs of Weyl points from the semi-Dirac points. In the presence of an additional inversion symmetry-breaking lattice distortion, the Weyl points acquire an energy gap, and an enhanced intrinsic anomalous Hall effect is realized. 
The presence of the Weyl points can produce negative magnetoresistance induced by chiral anomaly~\cite{Son_PRB2013}. The Berry phase at the semi-Dirac point is tunable by the octahedral rotation that can be controlled externally by sample thickness, or advanced interface engineering. Our findings serve as an important stepping stone towards understanding various kinds of quasiparticles that can be found in solids~\cite{Bradlyn_Sci2016,Wang_Nature2016,Wieder_Science2018,Tang_PRB2021,Yu_arXiv2021}.

\noindent {\bf Acknowledgements:} This work was supported by the U.S. Department of Energy, Office of Science, Basic Energy Sciences, Materials Sciences and Engineering Division.

\vspace{-3mm}
\def\bibsection{\section*{\refname}} 
\vspace{-3mm}

%

\newpage
\pagebreak
\clearpage
\widetext
\begin{center}
\textbf{{Supplemental Materials on "Semi-Dirac and Weyl Fermions in Transition Metal Oxides"}}
\vspace{1em}\\
{Narayan Mohanta$^{1}$, Jong Mok Ok$^{1}$, Jie Zhang$^{1}$, Hu Miao$^{1}$, Elbio Dagotto$^{1,2}$, Ho Nyung Lee$^{1}$, Satoshi Okamoto$^{1}$}\\
\it{$^{1}$Materials Science and Technology Division, Oak Ridge National Laboratory, Oak Ridge, TN 37831, USA}\\
\it{$^{2}$Department of Physics and Astronomy, The University of Tennessee, Knoxville, TN 37996, USA}

\end{center}
\setcounter{equation}{0}
\setcounter{figure}{0}
\setcounter{table}{0}
\setcounter{page}{1}
\makeatletter
\renewcommand{\theequation}{E\arabic{equation}}
\renewcommand{\thefigure}{S\arabic{figure}}
\renewcommand{\bibnumfmt}[1]{[R#1]}
\renewcommand{\citenumfont}[1]{R#1}

\vspace{5mm}

\noindent {\bf{\\I. Details of \textit{ab initio} calculations\\}} 

\noindent The electronic structure was computed within \textit{ab initio} density functional theory (DFT)~\cite{Hohenberg_PR1964} as implemented in the {\footnotesize QUANTUM ESPRESSO} code~\cite{Giannozzi_JPCM2009} with the plane-wave pseudopotential method. We adopted the Perdew-Burke-Ernzerhof generalized gradient approximation exchange-correlation functional~\cite{Perdew_PRL1996} and used a $16\times16\times16$  Monkhorst-Pack momentum-point mesh to discretize the first Brillouin zone. A plane wave cut-off of 600~eV, which were found to be sufficient to achieve convergence of the total energy. The energy convergence criterion was set to 10$^{-6}$~eV during the minimization process of the self-consistent cycle. Starting with the experimental lattice constants $a\!=\!5.518$~\AA, $b\!=\!5.518$~\AA, and $c\!=\!8.280$~\AA~\cite{JMOk_Unpublished}, full optimization of the crystal structure was performed using the force convergence criterion of $10^{-3}$~eV/\AA.  

The Bloch wave functions, obtained from the DFT calculations, were used to construct the maximally localized Wannier functions~\cite{Marzari_PRB1997,Souza_PRb2001,Mostofi_CompPhysCommun2008,Pizzi_JPCM2020}. Starting from an initial projection of atomic d-basis functions belonging to the $t_{2g}$ manifold and centered on metal sites, the $t_{2g}$-like Wannier functions were obtained. Using the tight-binding Hamiltonian, formulated using the $t_{2g}$ orbitals and obtained from Wannier90, the topological invariant was computed within the Wannier charge center formalism using the WannierTools code~\cite{Soluyanov_PRB2011,WU_CompPhysCommun2018}.\\

\noindent {\bf{\\II. Symmetry properties of the Dirac points in $I4/mcm$ space group\\}}

\begin{figure}[h]
\begin{center}
\includegraphics[width=0.5\columnwidth, clip]{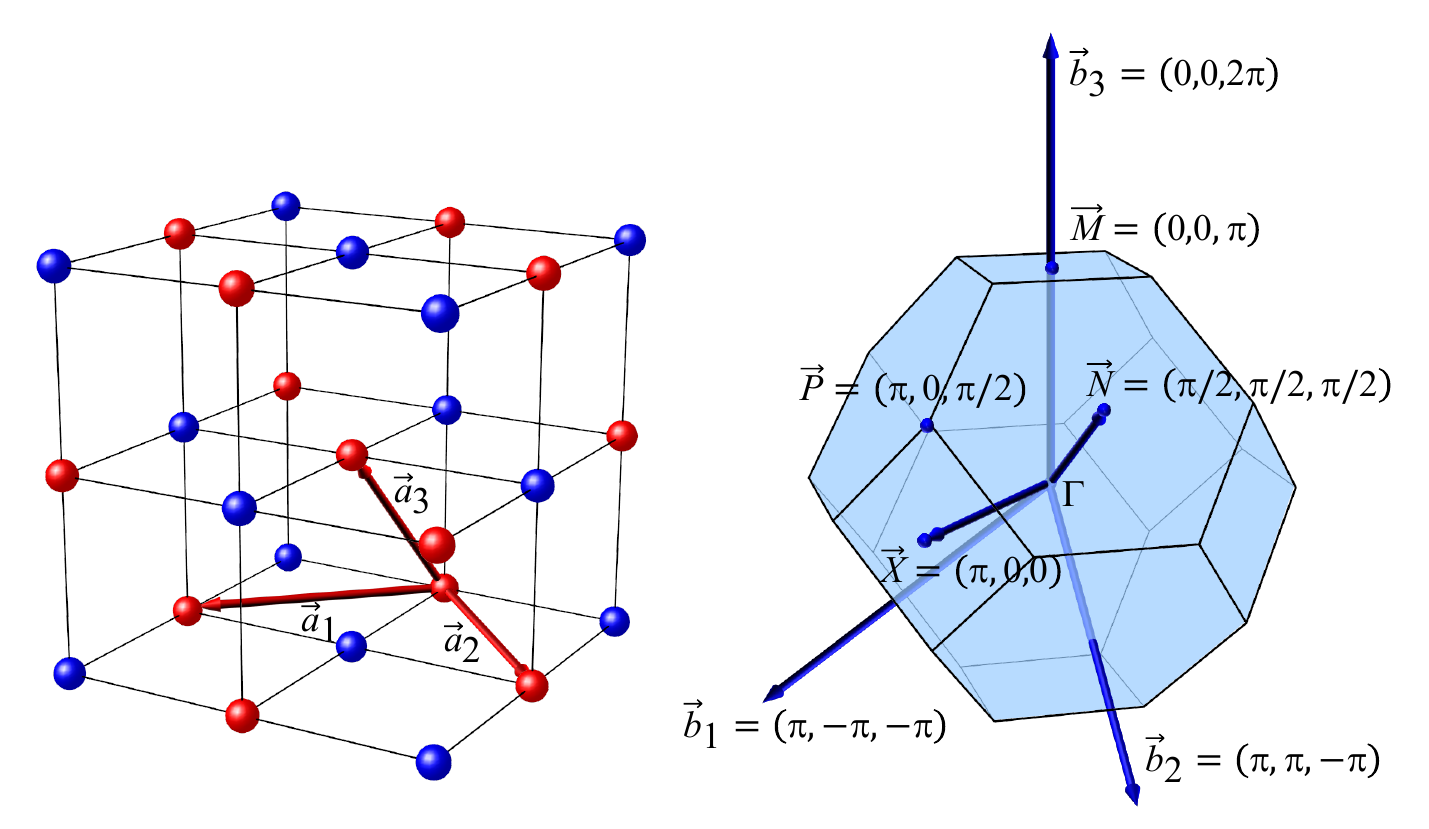}
\caption{Real and reciprocal lattices. 
Two sublattices are shown as read circles and blues circles on the left. Lattice translation vectors, $\vec a_{1,2,3}$, are indicated as red arrows. 
The first Brillouin zone is shown on the right, with three reciprocal lattice vectors, $\vec G_{12,3}$, and high symmetry points, $\Gamma$, X, N, M, and P, indicated by blue arrows. 
}
\label{fig:realandreciprocal}
\end{center}
\end{figure}

\noindent {\bf (a) Definition of lattice:}\\
For our discussion below, we first define real space and reciprocal space lattices. 
For the tetragonal lattice with space group 140, we consider a two-sublattice structure as shown in Fig.~\ref{fig:realandreciprocal} (left). 
Here, the distance between nearest-neighboring two sublattices is taken to be unity and, thus, 
lattice translation vectors are given by $\vec a_1=(1,-1,0)$, $\vec a_2=(1,1,0)$, and $\vec a_3=(1,0,1)$ in the Cartesian coordinate. 
With this notation, a half lattice translation connecting two sublattices is given by $(1,0,0)$, $(0,1,0)$ or $(0,0,1)$. 
Our coodinates are $-45$ degrees rotated about the $z$ axis 
from the standard notation with $\vec a_1=(a,0,0)$, $\vec a_2=(0,a,0)$, and $\vec a_3=(a/2,a/2,c)$ 
and half lattice translation given by $(a/2,a/2,0)$, $(a/2,-a/2,0)$, or $(0,0,c)$, which connects two sublattice sites. 
Here, the lattice constants $a$ and $c$ is shown explicitly. 

With our notation for $\vec a_{1,2,3}$, reciprocal lattice vectors become 
$\vec b_1= 2\pi (1/2,-1/2,-1/2)$,  $\vec b_2= 2\pi (1/2,1/2,-1/2)$, $\vec b_3= 2\pi(0,0,1)$ as shown in Fig.~\ref{fig:realandreciprocal} (right). 
High symmetry points are then given by 
$\vec X = 2\pi (1/2,0,0)$, 
$\vec M = 2\pi (0,0,1/2) $, 
$\vec N = 2\pi (1/4,1/4,1/4) $, and 
$\vec P = 2\pi(1/2,0,1/4)$. 
These high symmetry points to be used in the symmetry analysis are also indicated in Fig.~\ref{fig:realandreciprocal}. 
Using the standard notation, reciplocal lattice vectors and high symmetry points are given as follows:
$\vec b_1= 2\pi (1/a,0,-1/2c)$,  $\vec b_2= 2\pi (0, 1/a,-1/2c)$, $\vec b_3= 2\pi (0,0,1/c)$, 
$\vec X = 2\pi (1/2a,1/2a,0)$, $\vec M = 2\pi (0,0,1/2c) $, 
$\vec N = 2\pi (0,1/2a,1/4c) $, and 
$\vec P = 2\pi (1/2a,1/2a,1/4c)$.\\

\noindent {\bf (b) Space group operations:}\\
\noindent The Dirac points in the $I4/mcm$ (140) space group, appear at the high-symmetry momenta P and N. The non-symmorphic symmetry operations of this space group can be expressed using the Seitz notation $\{ \cal O| \mathbf{v}\}$, where $\cal O$ represents a point-group operation and  $\mathbf{v}\equiv(v_1,v_2,v_3)$
is a vector representing a translation~\cite{Bradley_Cracknell,Bradlyn_Sci2016,Elcoro_JAC2017}. The little group of a momentum $\mathbf k$ is the set of all space group operations whose ‘symmorphic part’ leaves the momentum $\mathbf k$ invariant up to an integer reciprocal lattice vector. The irreducible small co-representation of the little group and their corresponding matrix representations at the P point are given below

\begin{eqnarray}
E: \left[
\begin{array}{cccc}
1 & 0 & 0 & 0\\
0 & 1 & 0 & 0\\
0 & 0 & 1 & 0
\end{array}
\right]
: \tau^0 \sigma^0, &\hspace{1em}&
C_{2z}: \left[
\begin{array}{cccc}
-1 & 0 & 0 & 0\\
0 & -1 & 0 & 0\\
0 & 0 & 1 & 0
\end{array}
\right]
: \tau^0 (i \sigma^z), \nonumber \\
IC_{4z}: \left[
\begin{array}{cccc}
0 & 1 & 0 & 0\\
-1 & 0 & 0 & 0\\
0 & 0 & -1 & 0
\end{array}
\right]
: \tau^0 (1- i \sigma^z)/\sqrt{2}, &\hspace{1em}&
IC_{4 \bar z}: \left[
\begin{array}{cccc}
0 & -1 & 0 & 0\\
1 & 0 & 0 & 0\\
0 & 0 & -1 & 0
\end{array}
\right]
: \tau^0 (1+ i \sigma^z)/\sqrt{2}, \nonumber \\ 
C_{2 \bar y}|z: \left[
\begin{array}{cccc}
-1 & 0 & 0 & 0\\
0 & 1 & 0 & 0\\
0 & 0 & -1 & 1
\end{array}
\right]
: \tau^x ( i \sigma^y), &\hspace{1em}&
C_{2 \bar x}|z: \left[
\begin{array}{cccc}
1 & 0 & 0 & 0\\
0 & -1 & 0 & 0\\
0 & 0 & -1 & 1
\end{array}
\right]
: \tau^x ( i \sigma^x), \nonumber \\
M_{xy}|z: \left[
\begin{array}{cccc}
0 & -1 & 0 & 0\\
-1 & 0 & 0 & 0\\
0 & 0 &  1 & 1
\end{array}
\right]
: \tau^x ( i \sigma^x + i \sigma^y)/\sqrt{2}, &\hspace{1em}&
M_{x \bar y}|z: \left[
\begin{array}{cccc}
0 & 1 & 0 & 0\\
1 & 0 & 0 & 0\\
0 & 0 &  1 & 1
\end{array}
\right]
: \tau^x (  i \sigma^x - i \sigma^y)/\sqrt{2}. 
\end{eqnarray}

\noindent Here, the point-group operation $C_{n,x}$ indicates $n$-fold rotation about the $x$ axis. $E$ and $I$ represent the identity and the inversion operations, respectively, while  $M_{xy}$ represents the mirror operation with respect to the $x$-$y$ plane. $\sigma$ and $\tau$ are Pauli matrices acting on the spin and sublattice degrees of freedom, respectively. In the matrix representations, the first three columns show the point group operation $\cal O$, while the fourth column represents the translation $\mathbf{v}$.
The vector $(0,0,1)$ on the fourth columns indicates a half-translation along the $z$ direction; we use $1$ instead of $1/2$ because the nearest-neighbor bond connecting two sublattices is taken as the length unit.  Screw and glide operations indicated by $|z$ exchange sites at different sublattices, thus involve $\tau^x$. \\

\noindent {\bf (c) Symmetry enforced Dirac points:}\\
In this subsection, we discuss how the four-fold band degeneracy is protected by multiple glide symmetries at the P and the N points in SG-140. Here, we consider the half-translation explicitly by including $t_z=e^{i k_z}$, and use the following notation for the glide operations: $G_{xy, x \bar y}^z =t_z M_{xy, x \bar y}$. 

\subsubsection{Symmetry protected four-fold degeneracy at the N point}
\noindent We start from the protected degeneracy at the N point, which is one of the time reversal invariant momentums (TRIMs). The N point as well as the $\Gamma$ and the M points, are all TRIMs and in the glide mirror plane $M_{x \bar y}$.  Thus, we consider the glide operator $G_{x \bar y}^z =t_z M_{x \bar y}$. 
Following Ref. \cite{Wieder2016}, the commutation of $G_{x \bar y}^z$ and inversion $P$ at TRIMs is evaluated as 
\begin{eqnarray}
G_{x \bar y}^z P = t_z M_{x \bar y} P = P t_{-z} M_{x \bar y}= t_{2z} P t_z M_{x \bar y}=e^{2ik_z} P G_{x \bar y}^z. 
\end{eqnarray}
This implies that $\{G_{x \bar y}^z, P\}=0$ at the N point, where $k_z=\pi/2$ in the current unit, and $[G_{x \bar y}^z, P]=0$ at the $\Gamma$ and M points. Therefore, the states are fourfold degenerate at the N point and twofold degenerate at the $\Gamma$ and M points. Away from the N point, two states belonging to a local Kramers pair have different $G_{x \bar y}^z$ eigenvalues and, therefore, they can only anti-cross. 

\subsubsection{Symmetry protected four-fold degeneracy at the P point}
\noindent Next we consider the degeneracy at the P point. While this is not a TRIM, glide symmetries enforce the fourfold degeneracy, i.e. the Dirac point. Similar argument for space group 108 was presented in Ref.~\cite{Xia2020}. The generators of space group 140, that contain rotation symmetry $C_{2z}$ and four glide symmetries $G_{x,y}^z$ and $G_{xy, x \bar y}^z$ acting on the real space and the spin space, are 
\begin{eqnarray} 
C_{2z}: (x,y,z) &\rightarrow& (-x,-y,z) \otimes (i \sigma^z), \\
G_x^z = t_z M_x: (x,y,z) &\rightarrow& (-x,y,z+1) \otimes (i \sigma^x), \\
G_y^z = t_z M_y: (x,y,z) &\rightarrow& (x,-y,z+1) \otimes (i \sigma^y), \\
G_{xy}^z = t_z M_{xy}: (x,y,z) &\rightarrow& (-y,-x,z+1) \otimes \frac{i}{\sqrt{2}}(\sigma^x+\sigma^y), \\
G_{x \bar y}^z = t_z M_{x \bar y}: (x,y,z) &\rightarrow& (y,x,z+1) \otimes \frac{i}{\sqrt{2}} (\sigma^x - \sigma^y).  
\end{eqnarray}
The small group at the P point consists of $G_{x,y}^z$ and $C_{2z}$. 
Additionally, the P point is invariant under combined symmetries of 
$\Theta_{xy, x\bar y} \equiv \theta G_{xy, x \bar y}^z$, where $\theta = i \sigma^y K$ is the time reversal operator 
with $K$ the complex conjugation operator. 

Since the following relations hold at the P point: $(C_{2z})^2=(\Theta_{xy, x \bar y})^2=-1$ and $(G_{x, y}^z)^2=1$, the allowed eigenvalues of these operators are $\lambda (C_{2z}) = \lambda (\Theta_{xy, x \bar y})= \pm i$ and $\lambda (G_{xy, x \bar y}^z)= \pm 1$. 
The degenerate states at the P point are characterized by the combination of these symmetry operations. 

Now, it is straightforward to see the following anti-commutation relations \cite{Xia2020}  
$\{ G_{x,y}^z, C_{2z} \} = \{ \Theta_{xy, x \bar y}^z, C_{2z} \} = \{ G_{x}^z, G_y^z \}=0$ 
as well as the cross-commutation relations 
$\Theta_{xy} G_x^z = G_x^z \Theta_{x \bar y}$. 
These commutation relations allow one to construct four mutually orthogonal states. 
Assuming $|\psi_1 \rangle$ has eigenvalues of $\lambda (G_x^z)=1$ and $\lambda (C_{2z})=i$, other three states are constructed as 
\begin{eqnarray}
|\psi_2 \rangle &=& G_y^z | \psi_1 \rangle, \\
|\psi_3 \rangle &=& (\Theta_{xy}+\Theta_{x \bar y}) | \psi_1 \rangle, \\
|\psi_4 \rangle &=& (\Theta_{xy} - \Theta_{x \bar y}) | \psi_1 \rangle. 
\end{eqnarray}
Here, $|\psi_2 \rangle$ has eigenvalues $\lambda (G_x^z)=-1$ and $\lambda (C_{2z})=-i$, 
$|\psi_3 \rangle$ has $\lambda (G_x^z)=1$ and $\lambda (C_{2z})=-i$, 
$|\psi_2 \rangle$ has $\lambda (G_x^z)=-1$ and $\lambda (C_{2z})=-i$. 
Thus, these four states, at the P point, constitute the four-fold degenerate states, \textit{i.e.} the symmetry enforced Dirac point. 

To summarize this subsection, the symmetry arguments presented above are consistent with our density functional theory results, 
as well as our effective tight-binding model analyses and effective $\vec k \cdot \vec p$ model analysis, to be presented in the subsequent sections. 

\noindent {\bf{\\III. Effective tight-binding model\\}} 

\begin{figure}
\begin{center}
\includegraphics[width=0.7\columnwidth, clip]{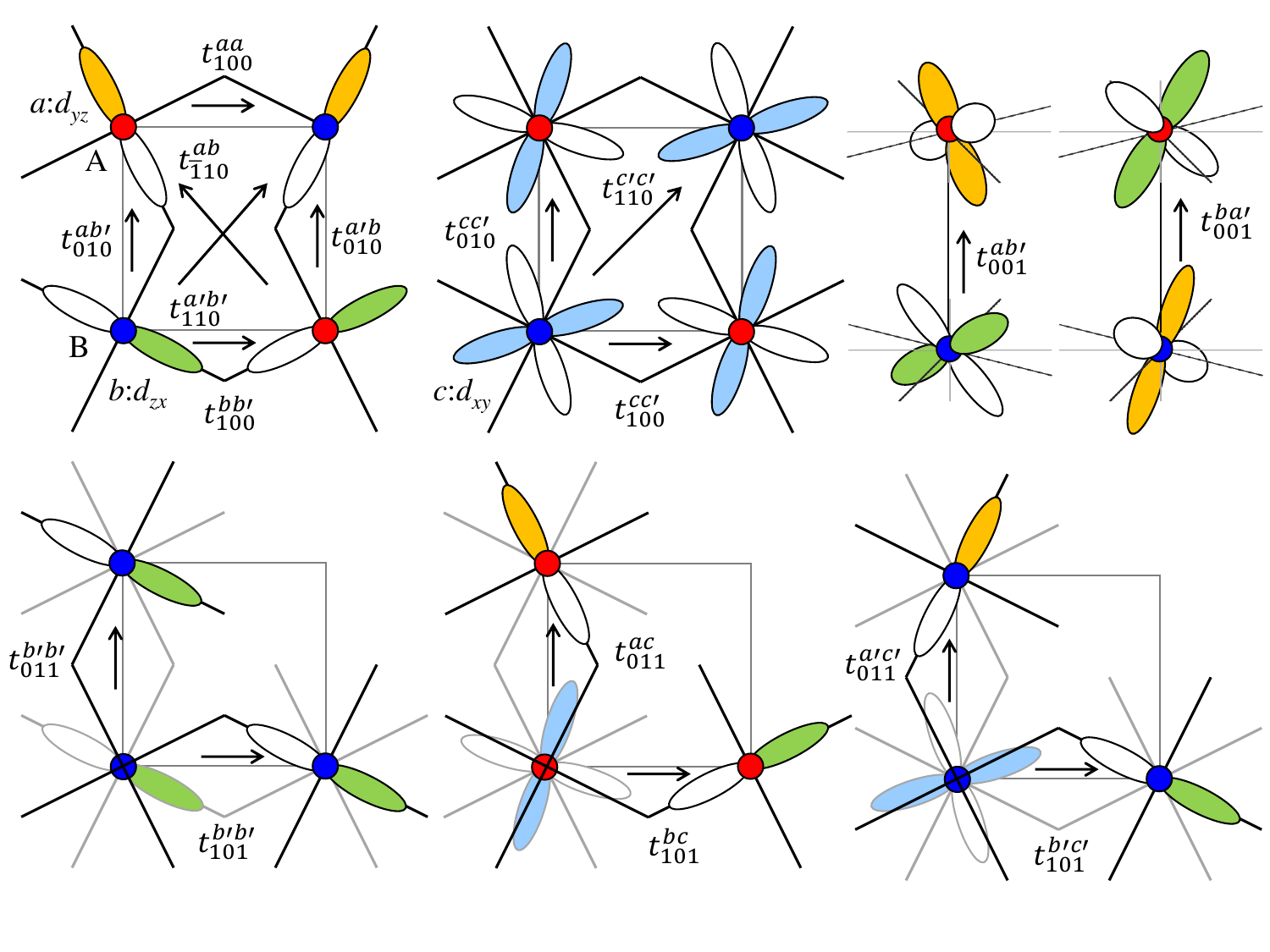}
\caption{Typical hopping integrals. Hopping integral  from orbital $\beta$ to orbital $\alpha$ along 
the  $h \hat x + k \hat y + l \hat z$ direction $t_{hkl}^{\alpha \beta}$ is indicated by an arrow. 
Here, red and blue circles belong to sublattice A and B, respectively. 
Orbitals $a:d_{yz}$, $b:d_{zx}$, and $c:d_{xy}$ are indicated by orange, green, blue robes, respectively. 
To highlight the sign difference in electron orbitals, negative regions are shown in white. 
Solid lines indicate how O$_6$ octahedra rotate, creating bond angles smaller than 180 degrees. 
In the second row, hopping is between second-neighbor sites on two adjacent layers. }
\label{fig:typicalhopping}
\end{center}
\end{figure}

\noindent To gain insight into the nature of the semi-Dirac dispersion in perovskite transition-metal oxides 
with the $I4/mcm$ crystalline symmetry, we develop an effective tight-binding Hamiltonian using a basis of the $t_{2g}$ orbitals. For the sake of simplicity, we define the local orbitals as $\{a,b,c\}=\{d_{xy}, d_{yz}, d_{zx} \}$, 
and denote the hopping integral from orbital $\beta$ to $\alpha$ along the $h \hat x + k \hat y + l \hat z$ direction as 
$t_{hkl}^{\alpha \beta}$. 
This leads to the following tight-binding Hamiltonian:
\begin{eqnarray}
H_{TB} =-\sum_{\alpha, \beta} \sum_{\vec r, \vec r'} \sum_\sigma 
 t_{\vec r-\vec r'}^{\alpha \beta} \alpha_{\vec r \sigma}^\dag \beta_{\vec r' \sigma}, 
\end{eqnarray}
where $\alpha_{\vec r \sigma}^{(\dag)}$ is an annihilation (creation) operator of an electron on $\alpha$ orbital at position $\vec r$ with spin $\sigma$. 
Typical hopping integrals are presented in Fig. \ref{fig:typicalhopping}. 
The amplitude and the sign of $t_{hkl}^{\alpha \beta}$ are determined by the relative direction $h \hat x + k \hat y + l \hat z$ and the sign of two orbitals 
and must  satisfy the underlying $I4/mcm$ symmetry. 
Here, the distance between the nearest-neighboring sites is taken to be unity. 
This leads to  the following relations:
\begin{eqnarray}
&t_{100}^{aa}=t_{\bar 100}^{aa}=t_{010}^{bb}=t_{0\bar 10}^{bb},& \nonumber \\
&t_{010}^{aa}=t_{0\bar 10}^{aa}=t_{100}^{bb}=t_{\bar 100}^{bb},& \nonumber \\
&t_{001}^{aa}=t_{00\bar 1}^{aa}=t_{001}^{bb}=t_{00\bar 1}^{bb},& \nonumber \\
&t_{100}^{a'b}=t_{\bar 100}^{a'b}=t_{010}^{a'b}=t_{0\bar 10}^{a'b}
=-t(\{a,b\} \leftrightarrow \{a',b'\}),& \nonumber \\
&t_{001}^{ab}=t_{00\bar 1}^{ab},& \nonumber \\
&t_{100}^{cc}=t_{\bar 100}^{cc}=t_{010}^{cc}=t_{0\bar 10}^{cc},& \nonumber \\
&t_{001}^{cc}=t_{00\bar 1}^{cc}& 
\end{eqnarray}
for nearest-neighor bonds, 
and 
\begin{eqnarray}
&t_{110}^{aa}=t_{\bar 1 \bar 10}^{aa}=t_{\bar 110}^{bb}=t_{1 \bar 10}^{bb}
=t_{110}^{b'b'}=t_{\bar 1 \bar 10}^{b'b'}=t_{\bar 110}^{a'a'}=t_{1 \bar 10}^{a'a'}
,& \nonumber \\
&t_{1 \bar 10}^{aa}=t_{\bar 110}^{aa}=t_{110}^{bb}=t_{\bar 1 \bar 10}^{bb}
=t_{1 \bar 10}^{b'b'}=t_{\bar 110}^{b'b'}=t_{110}^{a'a'}=t_{\bar 1 \bar 10}^{a'a'},& \nonumber \\
&t_{110}^{ab}=t_{\bar 1 \bar 10}^{ab}
=-t_{1 \bar 10}^{ab}=-t_{\bar 1 10}^{ab},& \nonumber \\
&t_{110}^{cc}=t_{\bar 1 \bar 10}^{cc}=t_{\bar 110}^{cc}=t_{1 \bar 10}^{cc},& \nonumber \\
%
&t_{011}^{aa}=t_{0\bar 1 \bar 1}^{aa}=t_{0 \bar 11}^{aa}=t_{0 1 \bar 1}^{aa}=
t_{101}^{bb}=t_{\bar 1 0 \bar 1}^{bb} = t_{\bar 101}^{bb}=t_{1 0 \bar 1}^{bb},& \nonumber \\ 
&t_{101}^{aa}=t_{\bar 1 0 \bar 1}^{aa}=t_{\bar 1 01}^{aa}=t_{1 \bar 0 1}^{aa}=
t_{011}^{bb}=t_{0\bar 1 \bar 1}^{bb} = t_{0\bar 11}^{bb}=t_{01 \bar 1}^{bb},& \nonumber \\ 
&t_{101}^{ac}=t_{\bar 1 0 \bar 1}^{ac} = - t_{\bar 101}^{ac}=- t_{1 0 \bar 1}^{ac}
=t_{011}^{bc}=t_{0 \bar 1  \bar 1}^{bc} = - t_{0 \bar 11}^{bc}=- t_{0 1 \bar 1}^{bc}, \nonumber \\
%
&t_{101}^{bc}=t_{\bar 10 \bar 1}^{bc}=t_{011}^{ac}=t_{0 \bar 1 \bar 1}^{ac} 
=- t_{1 0 \bar1}^{bc}=- t_{\bar 0 1}^{bc} = -t_{01 \bar 1}^{ac} = -t_{0 \bar 1 1}^{ac}
=-t(\{a,b,c\} \leftrightarrow \{a',b',c'\})
\end{eqnarray}
for second-neighbor bonds, and those not listed here are zero up to second-neighbor bonds.  
When the hopping integral depends explicitly on the sublattice A and B, 
we use $\{a',b',c' \}$ for sublattice B. 
Further-neighbor hopping integrals could be considered, but including these terms does not alter our discussion. 

In the undistorted perovskite structure, many hopping integrals disappear reflecting the symmetry 
(positive and negative sign of the $t_{2g}$ orbitals). 
Among such hopping integrals, non-zero value of $t_{101}^{ac}$ and symmetry related hopping 
as well as finite spin-orbit coupling
are found to lift the band degeneracy of the nodal lines along the P-N direction,  
creating the semi-Dirac dispersion at the P point and Dirac dispersions at the N point. 

In addition to the hopping integrals, crystal field splitting can exists between orbitals $a$, $b$, and $c$. Such a splitting can be described by the following Hamiltonian
\begin{eqnarray}
H_{\rm CFS} = \sum_\sigma \bigl\{ \varepsilon_{ab} (a_\sigma^\dag a_\sigma +b_\sigma^\dag b_\sigma) 
+ \varepsilon_c c_\sigma^\dag c_\sigma \bigr\}, 
\end{eqnarray}
where $a$ and $b$ remain degenerate in the $I4/mcm$ symmetry. 

\noindent Finally, we include the most important ingredient, the spin-orbit coupling in our description. 
Here, we consider the atomic-like form, given by  
\begin{eqnarray}
H_{\rm SOC} =\lambda \bigl( \vec l \cdot \vec s \bigr)_{t_{2g}} = \frac{\lambda}{2} \sum_{\alpha, \beta, \gamma} 
\sum_{\sigma, \sigma'}
i \varepsilon_{\alpha \beta \gamma}  \alpha_\sigma^\dag \sigma_{\sigma \sigma'}^\gamma \beta_{\sigma'}, 
\end{eqnarray}
where $(\ldots)_{t_{2g}}$ indicates the projection onto the $t_{2g}$ subspace; 
$\alpha$, $\beta$, and $\gamma$ are the orbital indices, 
$\sigma^\gamma$ represents the Pauli matrices, and 
$\varepsilon_{\alpha \beta \gamma}$ is the Levi-Civita antisymmetric tensor. 
For $H_{\rm CFS}$ and $H_{\rm SOC}$, site indices are omitted for simplicity. 

\noindent Thus, the effective Hamiltonian is given by the sum of the above three Hamiltonians as 
\begin{eqnarray}
H_{\rm eff} = H_{\rm TB} + H_{\rm CFS} + H_{\rm SOC}. 
\end{eqnarray}

\begin{figure}
\begin{center}
\includegraphics[width=0.9\columnwidth, clip]{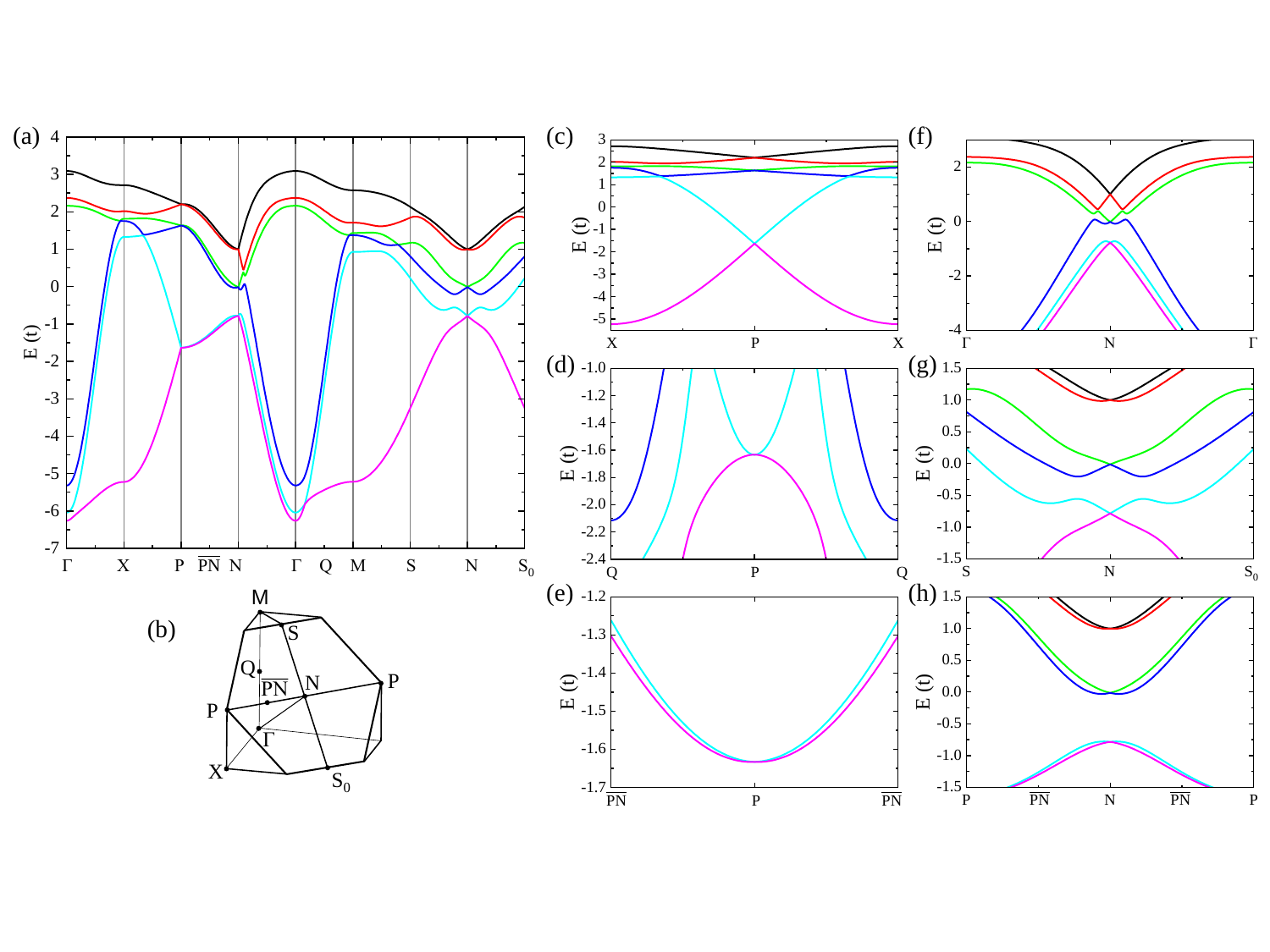}
\caption{Tight-binding dispersions. (a) Dispersions in the entire  Brillouin zone with its first octant shown in (b). 
Dispersions around the P point and the N point are shown in panels (c-e) and (f-h), respectively. 
The midpoint between P and N is labeled as $\overline{\rm PN}$. }
\label{fig:TB_summary}
\end{center}
\end{figure}

\noindent In an example that we show here, the following parameter set is considered:  
$t_{100}^{aa}=t_{001}^{aa}=t_{100}^{cc}=t$,
$t_{010}^{aa}=t_{001}^{cc}=0.1 t$, 
$t_{100}^{a'b} = 0.05 t$, 
$t_{001}^{a'b} = 0.1 t$,
$t_{110}^{cc} = t_{011}^{aa} = 0.4 t$, 
$t_{1\bar1 0}^{aa} =0.1 t$, 
$t_{110}^{ab} = 0.1 t$, 
$t_{101}^{ac} = 0.05 t$, 
$\varepsilon_c=0.2t$, 
$\lambda =0.5 t$, and others are set to zero.  
Resulting dispersion relation is summarized in Fig.~\ref{fig:TB_summary}. 
One can notice that this tight-binding dispersion relation qualitatively reproduces the DFT dispersions, 
in particular, the semi-Dirac dispersion at the P point and the linear Dirac dispersios at the N point. 

\begin{figure}
\begin{center}
\includegraphics[width=0.5\columnwidth, clip]{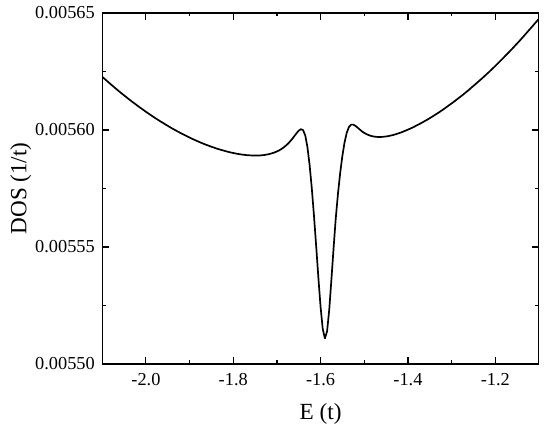}
\caption{Partial density of states originating from one of semi-Dirac dispersions at the P point. 
}
\label{fig:dosP}
\end{center}
\end{figure}

\noindent To see the consequence of the semi-Dirac dispersions, we computed the partial density of states (DOS) by focusing on one of the P point. 
The resulting DOS is shown in Fig. \ref{fig:dosP}. 
Here, the momentum $\vec k$ integral is carried out in a small cube around the P point with 
$\Delta k = \pm \pi/8$ and the Dirac delta function approximated as the Lorentzian with the width $\eta =0.02 t$. 
One notices a sharp dip in DOS, instead of a gap, reflecting the semi-Dirac dispersions. 
With correlation effects, this may lead to novel many-body effects as discussed in Ref. \cite{Wang2021}. 


\noindent {\bf{\\IV. Low-energy effective two-band model and $\vec k \cdot \vec p$ model\\}} 

\noindent The tight-binding model developed in the previous section is useful to further derive a low-energy effective model.  
For this purpose, we focus on the Kramers pairs that are eigenstates of the local Hamiltonian, $H_{\rm CFS}+H_{\rm SOC}$. 
Considering the fact that the eigenstates of $|l=1, l_z=\pm 1 \rangle$ are given by 
$|l=1, l_z=\pm 1 \rangle =| l_\pm \rangle = (\mp | a\rangle + i |b \rangle )/\sqrt{2}$, 
three sets of Kramers pairs can be expressed as 
\begin{eqnarray}
&|j_z= 3/2, \sigma \rangle = | l_{\sigma}, \sigma \rangle , & \\
&|\zeta, \sigma \rangle = u_\zeta  | l_{\bar \sigma}, \bar \sigma \rangle + v_\zeta | c, \sigma \rangle, &   \\
&|\eta, \sigma \rangle = u_\eta  | l_{\bar \sigma}, \bar \sigma \rangle + v_\eta | c, \sigma \rangle,  & 
\end{eqnarray} 
where the first Kramers pair has the total angular momentum $|l_z+s_z| = 3/2$, 
and the second  and the third Kramers pairs are obtained by diagonalizing $H_{\rm CFS}+H_{\rm SOC}$, with coefficients $u$ and $v$ being real. 
Now we take one of Kramers pairs and introduce the hopping integrals. 
Considering the hopping integrals described in the previous section, a low-energy Hamiltonian can be parameterized as  
\begin{eqnarray}
H_{low-E}&=& 
-2 A \cos k_x \cos k_y \tau_0 \sigma_0 
+\frac{1}{2} B \tau_2  (\sin 2 k_x  \sigma_1 - \sin 2 k_y \sigma_2) \sin k_z \nonumber \\
&&-2 \bigl\{C (\cos k_x + \cos k_y) + D \cos k_z  \bigr\} \tau_1 \sigma_0
+ 2 E  (\cos k_x + \cos k_y)  \tau_2 \sigma_3 ,
\label{eq:HlowE1}
\end{eqnarray}
where $\tau_i$ and $\sigma_j$ are $2\times2$ Pauli matrices acting on the sublattice space and the spin space specifying Kramers states, respectively, and Kronecker product are abbreviated as $\tau_i \otimes \sigma_j \equiv \tau_i \sigma_j$. 
The second term, involving $B$, represents the mixing between different Kramers states. 
This term is generated by hopping from the basis Kramers states, say $|\zeta, \sigma \rangle$, to 
other Kramers states, say $|\eta, \sigma \rangle $ or $|j_z= 3/2, \sigma \rangle $, and then to the basis Kramers states along 
$2 \hat x +1 \hat z$ and symmetry-related directions. 
Such processes are allowed by $t_{101}^{ac}$ and symmetry related hopping terms due to the tetragonal distortion.\\

\begin{figure}
\begin{center}
\includegraphics[width=0.6\columnwidth, clip]{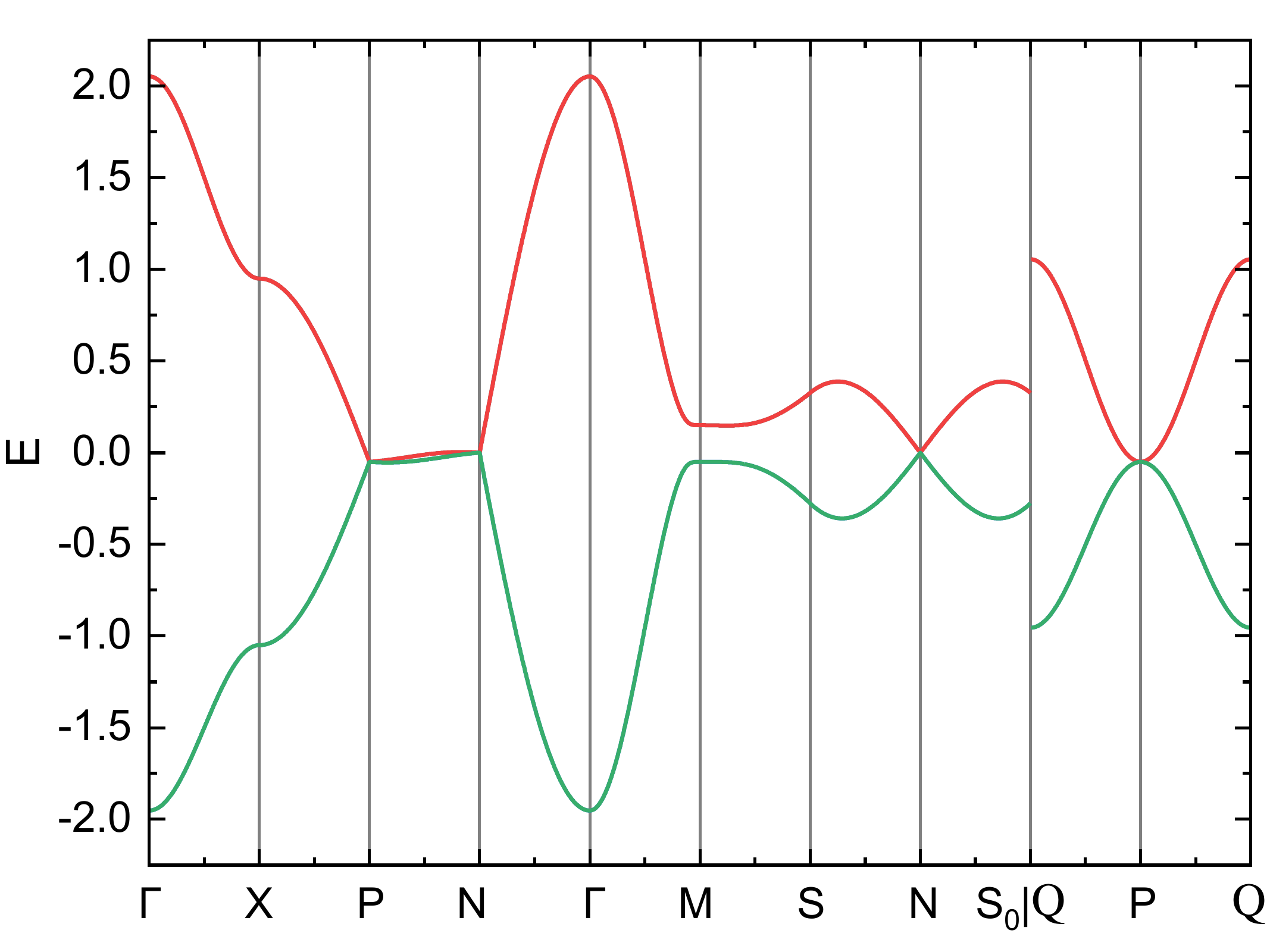}
\caption{Dispersion relation from the low-energy effective model Eq.~(\ref{eq:HlowE1}). 
Parameter values are chosen to be $A=0.05$, $B=0.02$, $C=0.5$, $D=1.0$, and $E=1.0$. }
\label{fig:dispersionTB}
\end{center}
\end{figure}

\noindent Figure~\ref{fig:dispersionTB} shows the dispersion relation from Eq. (\ref{eq:HlowE1}). 
One can notice that this dispersion reproduces the essential feature of the DFT band structure: 
1. semi-Dirac dispersion at the P point, 2. Dirac dispersion at the N point, and the gap opening at the M point. 

\noindent Now expanding cosine and sine functions near the P point, we arrive at 
\begin{eqnarray}
H_{low-E} (\vec P+\vec q)\approx 
-A \bigl( q_x^2 + q_y^2 \bigr) \tau_0 \sigma_0 
+B  \tau_2 ( q_x \sigma_1 - q_y  \sigma_2)  
-\Bigl\{ C \bigl( q_x^2 - q_y^2 \bigr) - D k_z  \Bigr\} \tau_1 \sigma_0 
+E \bigl( q_x^2 - q_y^2 \bigr) \tau_2 \sigma_3, 
\label{eq:HlowE2}
\end{eqnarray}
\noindent where a constant term is neglected from the diagonal component. 

\noindent The eigenvalues of this Hamiltonian are obtained as
\begin{eqnarray}
E_{\pm} (\vec P + \vec q) = -A \bigl( q_x^2 + q_y^2 \bigr) \pm \sqrt{B^2  \bigl(q_x^2 + q_y^2 \bigr) 
+ \Bigl\{ C \bigl( q_x^2 - q_y^2 \bigr) - D q_z \Bigr\}^2
+E^2 \bigl(q_x^2 - q_y^2 \bigr)^2
}. 
\end{eqnarray}

\begin{figure}
\begin{center}
\epsfig{file=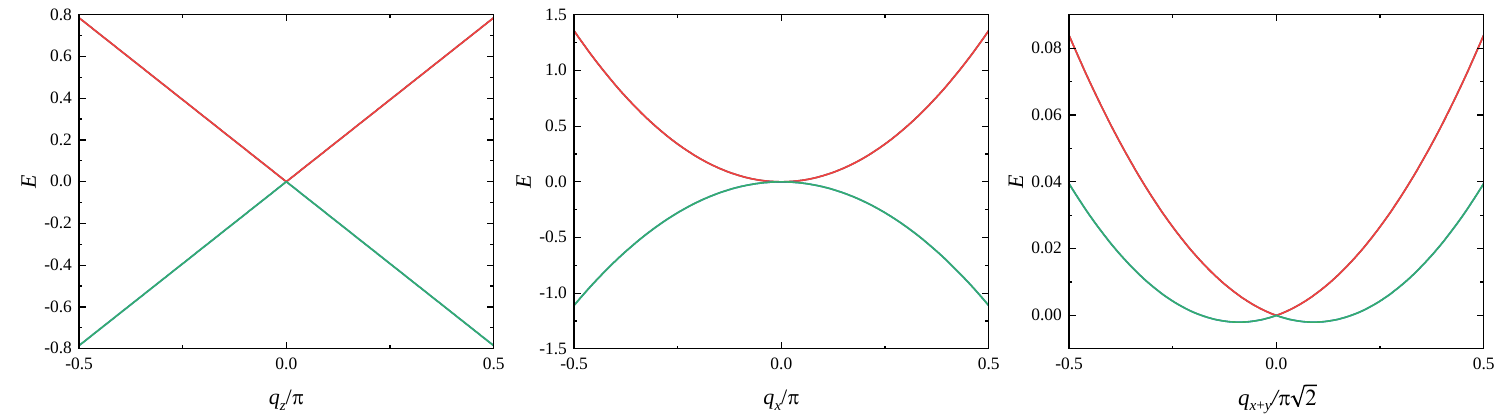,trim=0.0in 0.0in 0.0in 0.0in,clip=false, width=180mm}
\caption{Dispersion relation from the low-energy effective model Eq.~(\ref{eq:HlowE2}) along three inequivalent directions. 
Parameter values are chosen to be $A=-0.05$, $B=0.01$, $C=0.5$, and $D=1.0$. }
\label{fig:effective}
\end{center}
\end{figure}

\noindent In Fig. \ref{fig:effective}, we plot the dispersion relations along three different directions with 
$A=0.05$, $B=0.02$, $C=0.5$, $D=1.0$, and $E=1.0$. 
Linear band crossing along the $q_z$ direction, quadratic band crossing along the $q_x$ direction (and equivalently $q_y$), 
and semi-Dirac dispersion along the $\vec q_x \pm \vec q_y$ direction.  
The band splitting along the $\vec q_x \pm \vec q_y$ direction, corresponding to the P-N-P direction, 
is due to the parameter $B$.

\noindent In the undistorted perovskite structure, many hopping integrals disappear reflecting the symmetry 
(positive and negative sign of $t_{2g}$ orbitals). 
Among such hopping integrals, non-zero value of $t_{101}^{ac}$ and symmetry related hopping 
as well as finite spin-orbit coupling
are found to lift the band degeneracy of nodal lines along the P-N direction,  
creating semi-Dirac dispersions at the P point and Dirac dispersions at the N point. 

\noindent In addition to the hopping integrals, there could exist crystal field splitting between orbitals $a$, $b$, and $c$ as 
\begin{eqnarray}
H_{\rm CFS} = \sum_\sigma \bigl\{ \varepsilon_{ab} (a_\sigma^\dag a_\sigma +b_\sigma^\dag b_\sigma) 
+ \varepsilon_c c_\sigma^\dag c_\sigma \bigr\}, 
\end{eqnarray}
where $a$ and $b$ remain degenerate in the $I4/mcm$ symmetry. 

\noindent Finally, as the most important ingredient, the spin-orbit coupling is included. 
Here, we consider the atomic-like form given by  
\begin{eqnarray}
H_{\rm SOC} = \lambda \bigl( \vec l \cdot \vec s \bigr)_{t_{2g}} = \frac{\lambda}{2} \sum_{\alpha, \beta, \gamma} 
\sum_{\sigma, \sigma'}
i \varepsilon_{\alpha \beta \gamma}  \alpha_\sigma^\dag \sigma_{\sigma \sigma'}^\gamma \beta_{\sigma'}, 
\end{eqnarray}
where $(\ldots)_{t_{2g}}$ indicates the projection onto the $t_{2g}$ subspace, 
$\alpha$, $\beta$, and $\gamma$ are orbital indices, 
$\sigma^\gamma$ is the Pauli matrix, and 
$\varepsilon_{\alpha \beta \gamma}$ is the Levi-Civita antisymmetric tensor. 
For $H_{\rm CFS}$ and $H_{SOC}$, site indices are omitted for simplicity. 

\noindent Thus, the effective Hamiltonian is given by the sum of the above three terms:  
\begin{eqnarray}
H_{\rm eff} = H_{\rm TB} + H_{\rm CFS} + H_{\rm SOC}. 
\end{eqnarray}


\noindent In an example we show here, the following parameter set is considered:  
$t_{100}^{aa}=t_{001}^{aa}=t_{100}^{cc}=t$,
$t_{010}^{aa}=t_{001}^{cc}=0.1 t$, 
$t_{100}^{a'b} = 0.05 t$, 
$t_{001}^{a'b} = 0.1 t$,
$t_{110}^{cc} = t_{011}^{aa} = 0.4 t$, 
$t_{1\bar1 0}^{aa} =0.1 t$, 
$t_{110}^{ab} = 0.1 t$, 
$t_{101}^{ac} = 0.05 t$, 
$\varepsilon_c=0.2t$, 
$\lambda =0.5 t$, and others are set to zero.  
Resulting dispersion relation is summarized in Fig.~\ref{fig:TB_summary}. 
One notices that this tight-binding dispersion relation qualitatively reproduces the DFT dispersions, 
in particular semi-Dirac dispersions at the P point and the linear Dirac dispersion at the N point.\\ 

\noindent To see the consequence of the semi-Dirac dispersions, we computed the partial density of states (DOS) by focusing on one of the P points. 
The resulting DOS is shown in Fig. \ref{fig:dosP}. 
Here, the momentum $\vec k$ integral is carried out in a small cube around the P point with 
$\Delta k = \pm \pi/8$ and the Dirac delta function approximated as the Lorentzian with the width $\eta =0.02 t$. 
One notices a sharp dip in DOS, instead of a gap, reflecting the semi-Dirac dispersions. 
With correlation effects, this may lead to novel many-body effects as discussed in Ref. \cite{Wang2021}. 

\begin{figure}[t]
\begin{center}
\epsfig{file=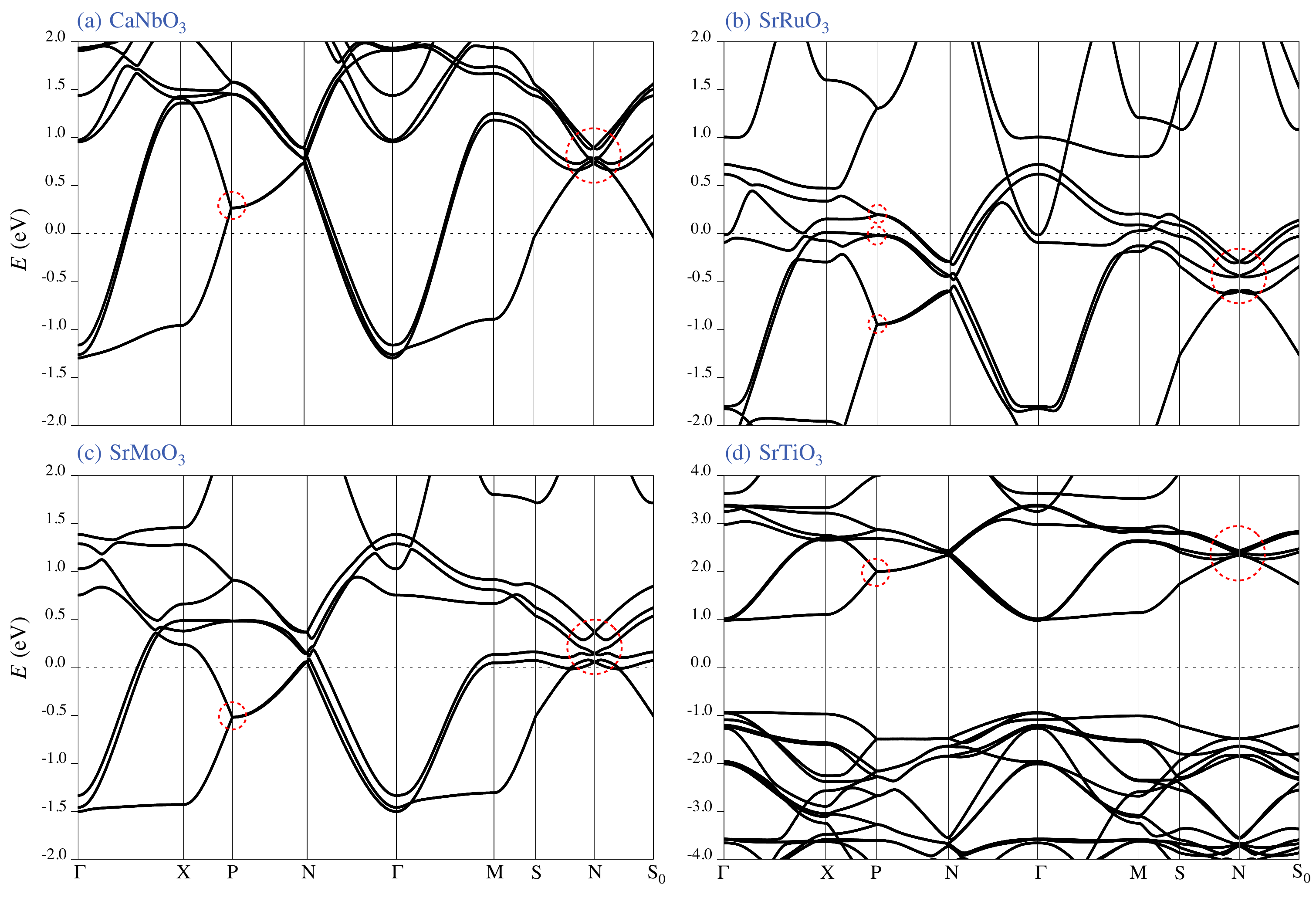,trim=0.0in 0.0in 0.0in 0.0in,clip=false, width=175mm}
\caption{DFT dispersions of tetragonal (a) CaNbO$_3$, (b) SrRuO$_3$, (c) SrMoO$_3$, and (d) SrTiO$_3$ in the presence of spin-orbit coupling. The red circles denote the three-dimensional semi-Dirac points at the P and the Dirac point at the N point. The lattice constants are: (a) $a=b=5.560$~\AA, $c=8.018$~\AA, (b) $a=b=5.577$~\AA, $c=8.087$~\AA, (c) $a=b=5.621$~\AA, $c=8.078$~\AA, (d) $a=b=5.567$~\AA, $c=7.907$~\AA.}
\label{figS2}
\vspace{-4mm}
\end{center}
\end{figure}
\begin{figure}[t]
\begin{center}
\epsfig{file=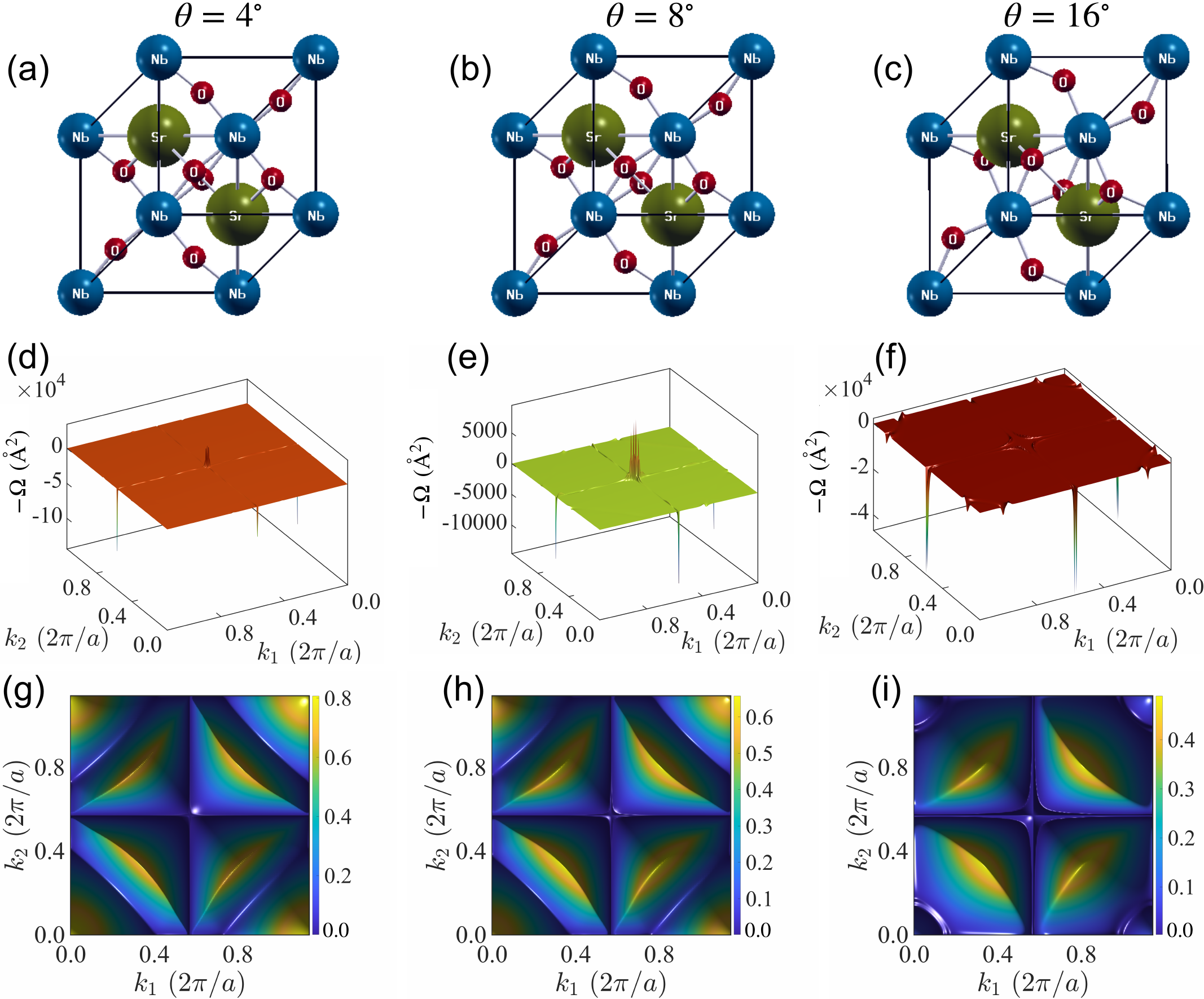,trim=0.0in 0.0in 0.0in 0.0in,clip=false, width=145mm}
\caption{Unit cell of tetragonal SrNbO$_3$ (top view) with different amount of the octahedral rotation: (a) $\theta \!=\!4^{\circ}$, (b) $\theta \!=\!8^{\circ}$, and (c) $\theta \!=\!16^{\circ}$. (d)-(f). Berry curvature at the respective values of $\theta$ with a Nb spin polarization $M\!=\!0.2$. (g)-(i) Energy gap (in eV) between the two bands that connect the semi-Dirac point for the above three values of $\theta$. Evidently, with increasing the rotation angle $\theta$, there are drastic changes in the Berry curvature $\Omega$, and the energy gap between the Bloch bands near the semi-Dirac points. }
\label{figS1}
\vspace{-5mm}
\end{center}
\end{figure}

\begin{figure}[t]
\begin{center}
\epsfig{file=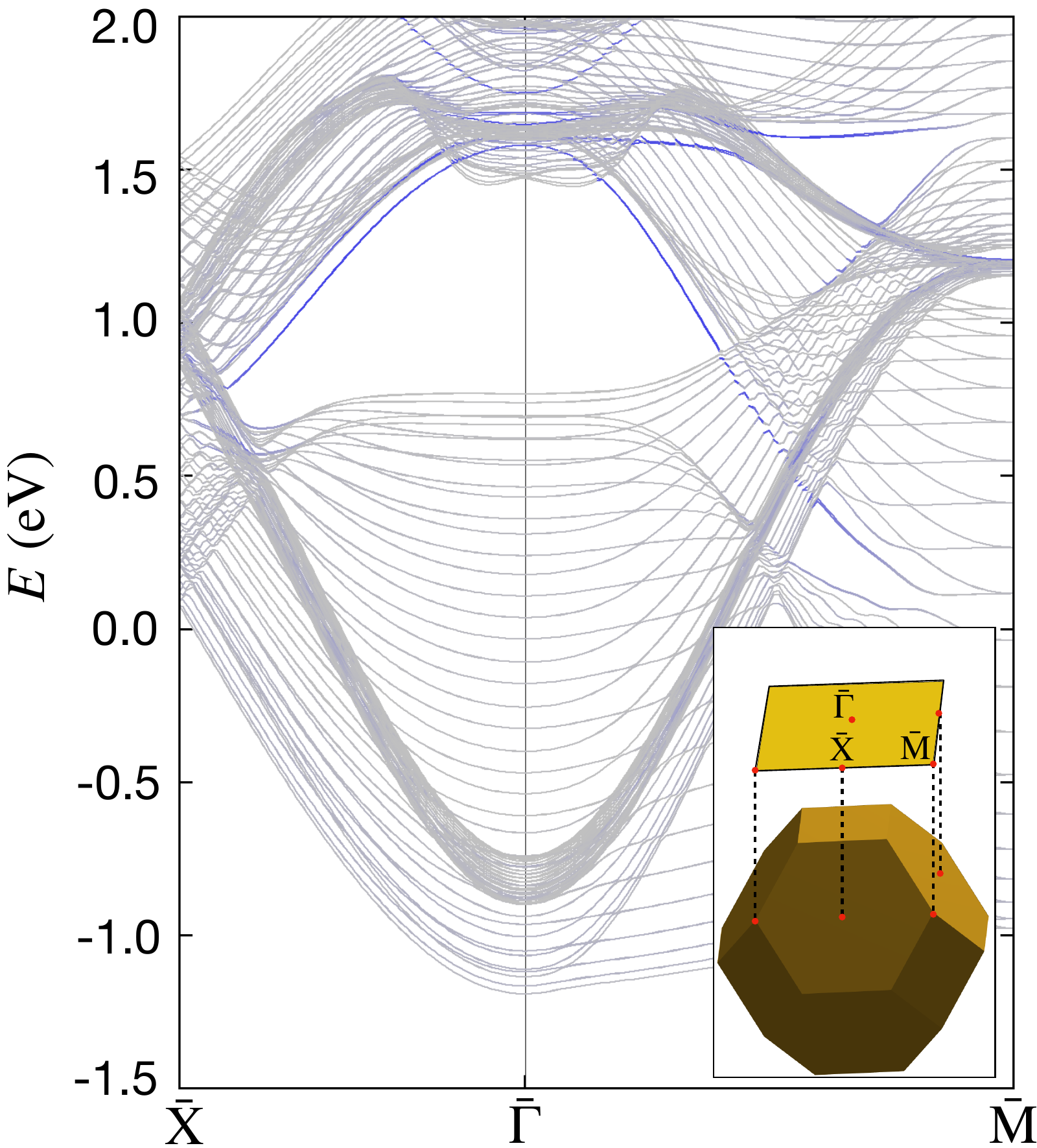,trim=0.0in 0.0in 0.0in 0.0in,clip=false, width=70mm}
\caption{Band dispersion of a (001) slab of tetragonal SrNbO$_3$, showing the surface states in blue and bulk states in grey along the high-symmetry momenta in the surface Brillouin zone, shown in the inset.  }
\label{figS4}
\vspace{-4mm}
\end{center}
\end{figure}

\noindent {\bf{\\V. Semi-Dirac points in the tetragonal oxides\\}} 

\noindent The presence of the 3D semi-Dirac fermions was investigated in a few other compounds from the \textit{I}4/\textit{mcm} space group, \textit{viz}, CaNbO$_3$, SrRuO$_3$, SrMoO$_3$, and SrTiO$_3$. In Fig.~\ref{figS2}, we show the electronic structures of these compounds that reveal the semi-Dirac points at the P point and Dirac points at the N point, similar to SrNbO$_3$. The semi-Dirac points in CaNbO$_3$ are located father away from the Fermi level than in  SrNbO$_3$. In SrRuO$_3$, a semi-Dirac point at the P point is located very close to the Fermi level while the Dirac points at the N point are below the Fermi level. Furthermore, SrRuO$_3$ is known to have magnetic order which will split the semi-Dirac points to Weyl points, as described in the main text. In SrMoO$_3$, the semi-Dirac point at the P point is located even farther away from the Fermi level while three Dirac points at the N point are located very close to the Fermi level. With slight electron doping, these semi-Dirac points can appear at the Fermi level and contribute to the transport properties. Therefore, SrMoO$_3$ is another candidate material that can be interesting to look at to investigate the signatures of the semi-Dirac points. On the other hand, SrTiO$_3$ has the semi-Dirac points $\sim$2~eV above the Fermi level and hence they will not influence the transport properties. SrTiO$_3$, is therefore, useful as a substrate compound to control the epitaxial strain in the metallic tetragonal oxides.

\noindent {\bf{\\VI. Octahedral rotation as a control parameter for the Berry curvature\\}} 

\noindent The electronic properties of tetragonal SrNbO$_3$ were studied at several values of the rotation angle $\theta$ of NbO$_6$ octahedra. In experiments, $\theta$ can be tuned by controlling the epitaxial strain, via an effective interface engineering, changing the sample thickness, or  an external electric field. In Fig.~\ref{figS1}, the results at three values $\theta=4^{\circ}$, $\theta=8^{\circ}$ and $\theta=16^{\circ}$ are shown at a finite Nb spin polarization $M=0.2$.  The Berry curvature $\Omega$ exhibits peaks at the P point (center of the plane in (d)-(f)) and the N points (side corners of the plane) --- the high-symmetry momenta at the Brilluoin zone boundary at which the semi-Dirac points appear. At the P point, $\Omega$ increases with increasing $\theta$ up to  $\theta \approx10^{\circ}$, after which it decreases. On the other hand, $\Omega$ at the N points increases monotonically with increasing $\theta$ up to  $\theta =16^{\circ}$. The energy separation between the two relevant bands near the semi-Dirac point, shown in Fig.~\ref{figS1}(g)-(i) at the three values of $\theta$, also captures the substantial changes in the band dispersion. The change in the Berry curvature will consequently be noticeable in the intrinsic anomalous Hall conductivity $\sigma_{xy}$. The octahedral rotation, therefore, provides an additional control knob to tune the Berry phase properties in oxides. 

\noindent {\bf{\\VII. Surface states in a slab (001) geometry\\}} 

\noindent To study the nature of the surface states of tetragonal SrNbO$_3$, we calculate the band dispersion in a slab geometry. We consider a slab of thickness 40 unit cells oriented along the (001) direction, relevant to the common experimental situation, and use the tight-binding Hamiltonian, expressed in the basis of the Wannier orbitals. The band dispersion of the SrNbO$_3$ slab, along the high-symmetry momenta in the surface Brillouin zone, is shown in Fig.~\ref{figS4}. In this case, the surface states do not exhibit any crossing-like feature, as expected. With a finite magnetization at the Nb site also, the surface states do not reveal any topological band crossing, as confirmed by the non-quantized anomalous Hall conductivity, as presented in the main text.

%

\end{document}